\documentclass[12pt]{article}
\usepackage{a4p,cite,rotating}
\newcommand{\z}{\mbox{$\rm Z^0$}}

\newcommand{\ccbar}{\mbox{$\rm c\overline{c}$}}

\newcommand{\ztobb}{\mbox{$\z\rightarrow {\rm b\overline{b}}$}}
\newcommand{\ztocc}{\mbox{$\z\rightarrow\ccbar$}}

\newcommand{\GeVc} {{\mathrm{GeV}}\!/c}
\newcommand{\GeVcc}{{\mathrm{GeV}}\!/c^2}

\newcommand{\MeVc} {{\mathrm{MeV}}\!/c}
\newcommand{\MeVcc}{{\mathrm{MeV}}\!/c^2}

\newcommand{\epluseminus}{\mbox{$\rm e^+e^-$}}

\newcommand{\meanxe}{\mbox{$\langle x_E\rangle$}}
\newcommand{\PLB}[3] {Phys.~Lett.\ {B#1} (#2) #3}
\newcommand{\PRL}[3] {Phys.~Rev.\ {Lett.~#1} (#2) #3}
\newcommand{\PRD}[3] {Phys.~Rev.\ {D#1} (#2) #3}
\newcommand{\NIM}[3] {Nucl.~Instr.\ {Meth.~#1} (#2) #3}

\newcommand{\CPC}[3] {Comp.~Phys.\ {Comm.~#1} (#2) #3}
\newcommand{\ZPC}[3] {Z.~Phys.\ {C#1} (#2) #3}

\newcommand{\EPJ}[3] {Eur.~Phys.~J.\ {C#1} (#2) #3}
\newcommand{\etal} {et~al.}
\newcommand{\bstar}{\mbox{${\rm B}^{*}$}}
\newcommand{\Bnull}{\mbox{${\rm B}_{0}^{*}$}}
\newcommand{\Bone}{\mbox{${\rm B}_{1}$}}
\newcommand{\Bna}{\mbox{${\rm B}_{1}(3/2)$}}
\newcommand{\Bbr}{\mbox{${\rm B}_{1}(1/2)$}}
\newcommand{\Btwo}{\mbox{${\rm B}_{2}^{*}$}}

\newcommand{\bstarj}{\mbox{${\rm B}^{*}_{J}$}}
\newcommand{\bstarjtob}{\mbox{$\bstarj \rightarrow {\rm B}\pi^{\pm}$}}
\newcommand{\bstarjtobs}{\mbox{$\bstarj \rightarrow {\rm B}^* \pi^{\pm}$}}
\newcommand{\bstarjtobb}{\mbox{$\bstarj \rightarrow {\rm B}\pi^{\pm}(X)$}}
\newcommand{\bstarjtobsb}{\mbox{$\bstarj \rightarrow {\rm B}^* \pi^{\pm}(X)$}}
\newcommand{\bstartob}{\mbox{$\bstar \rightarrow {\rm B}\gamma$}}
\newcommand{\bstarjs}{\mathrm{B^{*}_{\it sJ}}}

\newcommand{\brs}{\mbox{${\rm BR}\:\!(\;\!\bstarj \rightarrow {\rm B}^{*} \pi) $}}
\newcommand{\brsb}{\mbox{${\rm BR}\:\!(\;\!\bstarj \rightarrow {\rm B}^{*} \pi (X)) $}}

\newcommand{\dmbstar}{\mbox{$\Delta M$}}
\newcommand{\bsprob}{\mbox{${\cal W}({\rm B}^*) $}}
\newcommand{\gee}{\mbox{$\gamma \rightarrow {\rm e}^+{\rm e}^-$}}
\newcommand{\bpi}{\mbox{${\rm B}\pi$}}
\newcommand{\bpipm}{\mbox{${\rm B}\pi^\pm$}}
\newcommand{\wrt}{with respect to}
\newcommand{\brpipi}{\mbox{${\rm BR}\:\!(\;\!\bstarj \rightarrow {\rm B}^{(*)} \pi\pi) $}}
\newcommand{\bstarpipi}{\mbox{${\rm B}^{*} \pi\pi$}}
\newcommand{\bpipi}{\mbox{${\rm B \pi\pi}$}}

\input epsf
\newcommand{\epostfig}[3]{
\begin{figure}[tbp]
\setlength{\epsfxsize}{1.1\hsize}
\hspace*{-0.05\hsize} \epsfbox{./#1}
\caption{\label{#2}#3}
\end{figure}
}
\newcommand{\epostfigtwo}[3]{
\begin{figure}[tbp]
\vspace{1.0cm}
\setlength{\epsfxsize}{0.65\hsize}
\hspace*{+0.15\hsize} \epsfbox{./#1}
\vspace{1.5cm}
\caption{\label{#2}#3}
\end{figure}
}
\newcommand{\nhad}{4}
\newcommand{\lbevcut}{0.6}
\newcommand{\bpurity}{96}
\newcommand{\beffy}{49}
\newcommand{\brval}{0.85}
\newcommand{\brstat}{0.13}
\newcommand{\brstatplus}{+0.26}
\newcommand{\brstatminus}{-0.27}
\newcommand{\brsystot}{0.12}
\newcommand{\None} {8782}
\newcommand{\dNone}{252}
\newcommand{\Ntwo} {12051}
\newcommand{\dNtwo}{295}
\newcommand{\eff}     {0.578}
\newcommand{\deff}    {0.005}
\newcommand{\effnull} {1.566}
\newcommand{\deffnull}{0.010}
\newcommand{\effstar} {1.314}
\newcommand{\deffstar}{0.008}
\newcommand{\ese} {0.05084}
\newcommand{\dese}{0.00023}
\newcommand{\esd} {0.06680}
\newcommand{\desd}{0.00026}
\newcommand{\ee}  {0.04601}
\newcommand{\dee} {0.00030}
\newcommand{\ed}  {0.07962}
\newcommand{\ded} {0.00039}
\newcommand{\mb}{5.738}
\newcommand{\dmbplus}{0.005}
\newcommand{\dmbminus}{0.006}
\newcommand{\dmbsys}{0.007}
\newcommand{\gb}{18}
\newcommand{\dgbstatplus}{15}
\newcommand{\dgbstatminus}{13}
\newcommand{\dgbsysplus}{29}
\newcommand{\dgbsysminus}{23}
\newcommand{\bra}{0.74}
\newcommand{\dbrastatplus}{0.12}
\newcommand{\dbrastatminus}{0.10}
\newcommand{\dbrasysplus}{0.21}
\newcommand{\dbrasysminus}{0.15}
\begin{document}
\begin{titlepage}
\smallskip
\begin{center}
{\large
EUROPEAN ORGANIZATION FOR NUCLEAR RESEARCH}
\end{center}
\smallskip
{\flushright CERN-EP/2000-125\\September 19, 2000\\}
\bigskip
\begin{center}
  {\LARGE\bf\boldmath
    Investigation of the Decay of Orbitally-Excited
    \vspace{1.5mm}
    B Mesons and First Measurement of the
    Branching Ratio $\brsb$
  }
\end{center}
\bigskip
  
  \begin{center}

     \Large {\bf The OPAL Collaboration}

     \bigskip

    
    
    
\bigskip
    
  \end{center}
  
  \bigskip
\begin{abstract}
From about $\nhad$ million hadronic $\z$ decays recorded by the OPAL
detector on and near to the $\z$ resonance, we select a sample of more
than $570\, 000$ inclusively reconstructed B mesons.
Orbitally-excited mesons $\bstarj$ are reconstructed using $\bpipm$
combinations. Independently, $\bstar$ mesons are reconstructed using
the decay $\bstartob$. The selected $\bstar$ candidates are used to
obtain samples enriched or depleted in the decay $\bstarjtobsb$, where
$(X)$ refers to decay modes with or without additional
accompanying decay particles. From the number of signal candidates in
the $\bpipm$ mass spectra of these two samples, we perform the first
measurement of the branching ratio of orbitally-excited B mesons
decaying into $\bstar\pi(X)$:
\[
 \brsb = \brval^{\,\brstatplus}_{\,\brstatminus} \pm \brsystot \, ,
\]
where the first error is statistical and the second systematic.
If $\bstarj$ decay modes other than single pion transitions
can be neglected the measured ratio corresponds to the branching ratio
$\brs$.

In the framework of Heavy Quark Symmetry, a simultaneous fit to the
$\bpipm$ mass spectra of the samples enriched or depleted in
$\bstarjtobsb$ decays yields the mass and the width of the $\Bna$
state, as well as the branching ratio of $\bstarj$ mesons decaying
into $\bstar\pi$:
\begin{eqnarray*}
 M(\Bna)      & = & (\:\mb ^{\:+\:\dmbplus                      }_{\:-\:\dmbminus}\pm \dmbsys \:)          \;\GeVcc \\
 \Gamma(\Bna) & = & (\:\gb ^{\:+\:\dgbstatplus\:+\:\dgbsysplus  }_{\:-\:\dgbstatminus\:-\:\dgbsysminus}\:) \;\MeVcc \\
 \brs         & = &  \bra  ^{\:+\:\dbrastatplus\:+\:\dbrasysplus}_{\:-\:\dbrastatminus\:-\:\dbrasysminus}\; ,
\end{eqnarray*}
where the uncertainties are statistical and systematic, respectively.

\end{abstract}

\vspace{0.6cm}

\begin{center}
  (Submitted to Eur. Phys. J.)\\

%
\end{center}

\end{titlepage}
\begin{center}{\Large        The OPAL Collaboration
}\end{center}\bigskip
\begin{center}{
G.\thinspace Abbiendi$^{  2}$,
K.\thinspace Ackerstaff$^{  8}$,
C.\thinspace Ainsley$^{  5}$,
P.F.\thinspace {\AA}kesson$^{  3}$,
G.\thinspace Alexander$^{ 22}$,
J.\thinspace Allison$^{ 16}$,
K.J.\thinspace Anderson$^{  9}$,
S.\thinspace Arcelli$^{ 17}$,
S.\thinspace Asai$^{ 23}$,
S.F.\thinspace Ashby$^{  1}$
% ...
}\end{center}\bigskip
\bigskip
$^{  1}$School of Physics and Astronomy, University of Birmingham,
Birmingham B15 2TT, UK
\newline
$^{  2}$Dipartimento di Fisica dell' Universit\`a di Bologna and INFN,
I-40126 Bologna, Italy
\newline
$^{  3}$Physikalisches Institut, Universit\"at Bonn,
D-53115 Bonn, Germany
\newline
% ...
$^{  5}$Cavendish Laboratory, Cambridge CB3 0HE, UK
\newline
% ...
$^{  8}$CERN, European Organisation for Nuclear Research,
CH-1211 Geneva 23, Switzerland
\newline
$^{  9}$Enrico Fermi Institute and Department of Physics,
University of Chicago, Chicago IL 60637, USA
\newline
% ...
$^{ 16}$Department of Physics, Schuster Laboratory, The University,
Manchester M13 9PL, UK
\newline
$^{ 17}$Department of Physics, University of Maryland,
College Park, MD 20742, USA
\newline
% ...
$^{ 22}$Department of Physics and Astronomy, Tel Aviv University,
Tel Aviv 69978, Israel
\newline
$^{ 23}$International Centre for Elementary Particle Physics and
Department of Physics, University of Tokyo, Tokyo 113-0033, and
Kobe University, Kobe 657-8501, Japan
\newline
% ...
%
\eject
%
%
\section{Introduction}\label{s:intro}
 
An important prediction of Heavy Quark Effective Theory (HQET) is the
existence of an approximate spin-flavour symmetry for hadrons
containing one heavy quark $Q$ ($m_Q \gg \Lambda_{\rm QCD
}$)~\cite{hqet}. In the limit $m_Q \rightarrow \infty$, mesons
composed of a heavy quark $Q$ and a light quark $q$ are characterised
by the spin of the heavy quark, $S_Q$, the total angular momentum of
the light quark, $j_q = S_q + L$, and the total angular momentum, $J$,
where $S_q$ and $L$ denote spin and orbital angular momentum of
the light quark, respectively. In the heavy quark limit, both $S_Q$ and
$j_q$ are good quantum numbers and the total angular momentum of the
meson is given by $J = S_Q + j_q$. For $L=1$, there are four states
with spin-parity $J^P = 0^+$, $1^+$, $1^+$ and $2^+$. If the heavy
quark $Q$ is a bottom quark, these states are labelled $\Bnull$,
$\Bone$ for both $1^+$ states\footnote{In the limit $m_Q \rightarrow
\infty$, the notations $\Bbr$ and $\Bna$ are used. In the case of
mixing of the $J=1$ states, the notations ${\rm B}_1(H)$ and ${\rm
B}_1(L)$ are used to distinguish the physical states.}  and $\Btwo$
\cite{pdg}, respectively. The four states, commonly called $\bstarj$,
or alternatively ${\rm B}^{**}$
\footnote{Throughout this paper, we use the Particle Data Group
notation \bstarj\ for orbitally-excited B mesons.}, are grouped into
two sets of degenerate doublets, corresponding to $j_q=1/2$ and
$j_q=3/2$ as indicated in Table~\ref{t:states}. Parity and angular
momentum conservation put restrictions on the strong decays of these
states to $\rm B^{(*)}\pi$~\footnote{Throughout this paper, $\rm
B^{(*)}\pi$ denotes the final states $\bpi$ {\it and} $\bstar\pi$. The
notations $\rm B^{(*)}\pi\pi$ and $\bstar\pi (X)$ are to be
interpreted in the same way.} (see Figure~\ref{f:transition}). The
$0^+$ state can only decay to $\bpi$ via an S-wave transition, the
$1^+_{1/2}$ to $\bstar\pi$ via an S-wave transition, the $1^+_{3/2}$ to
$\bstar\pi$ via a D-wave transition, and the $2^+$ state can decay to
both $\bpi$ and $\bstar\pi$ via D-wave transitions only. States
decaying via an S-wave transition are expected to be much broader than
the states decaying via a D-wave transition\cite{godfreykokoski}. In
addition to the single pion transitions, decays to $\bstarpipi$
and $\bpipi$ are also possible. In the case of di-pion transitions,
all four $\bstarj$ states are allowed to decay to $\bstar$ as well as
to~B. Although these decays are phase-space suppressed, intermediate
states with large width like $\bstarj \rightarrow {\rm B}^{(*)}\rho
\rightarrow {\rm B}^{(*)} \pi \pi$ may cause a significant enhancement
of the $\bstarpipi$ and $\bpipi$ final states~\cite{ehq}. Additional
$\bstarj$ decay modes with other than one or two accompanying pions
are expected to be strongly suppressed but can not be
excluded. Therefore, the notations $\bstar\pi (X)$ and $\rm B\pi (X)$
are chosen to refer to the final states of $\bstarj$ decays.

Given the HQET predictions listed in Table~\ref{t:states}, the four
$\bstarj$ states are expected to overlap in mass. So far, in analyses
from LEP experiments
~\cite{bdstar,bdstardelphi,bdsalephexcl,bdsalephincl} 
and from CDF~\cite{bdscdf} $\bstarj$ mesons are reconstructed
in the ${\rm B}\pi$ final state only, observing one single peak in the
${\rm B}\pi$ mass spectrum. This is not sufficient to resolve any
substructure of the four expected $\bstarj$ states. In addition, for
decays to $\bstar\pi$ where the photon in the decay $\bstartob$ is not
detected, the reconstructed ${\rm B}\pi$ mass is shifted by
$M_{\rm B}-M_{\rm B^*}=-46\;\MeVcc$. A recent analysis\cite{bdstarl3} tries to cope with
these problems by constraining all properties of the four $\bstarj$
states according to HQET predictions except for the masses and widths
of $\Bbr$ and $\Btwo$.
\begin{table}[t]
\begin{center}
\begin{tabular}{|c|c||c|c|l||c|c|}

\hline
         &                &
\multicolumn{3}{c||}{predicted properties \cite{ebertfaustovgalkin,ehq,godfreykokoski}} &
\multicolumn{2}{c|}{Monte Carlo input}                                                  \\ \cline{3-7}
\raisebox{1.5ex}[-1.5ex]{state}    & \raisebox{1.5ex}[-1.5ex]{$J_j^P$}                  &
\small mass $\left[ {\GeVcc} \right]\!$                                                 &
\small width $\left[ {\GeVcc} \right]\!$                                                &
decay mode                                                                              &
\small mass $\left[ {\GeVcc} \right]\!$                                                 &
\small width $\left[ {\GeVcc} \right]\!$                                                \\
\hline
$\Bnull$ &  $0_{1/2}^{+}$ &  5.738 & 0.20-1.00 & $({\rm B}\pi)_{\rm S-wave}$  & 5.750 & 0.300  \\
$\Bone$  &  $1_{1/2}^{+}$ &  5.757 & 0.25-1.30 & $(\bstar \pi)_{\rm S-wave}$  & 5.770 & 0.300  \\
$\Bone$  &  $1_{3/2}^{+}$ &  5.719 & 0.021     & $(\bstar \pi)_{\rm D-wave}$  & 5.725 & 0.020  \\
$\Btwo$  &  $2_{3/2}^{+}$ &  5.733 & 0.025     & $(\bstar \pi)_{\rm D-wave}$, & 5.737 & 0.025  \\
         &                &        &           & $({\rm B} \pi)_{\rm D-wave}$ &       &        \\
\hline
\end{tabular}
\caption{\label{t:states} Masses, widths and dominant decay modes
based on theoretical predictions
\cite{ehq,ebertfaustovgalkin,godfreykokoski,manyothers,isgur} and the
corresponding Monte Carlo input values used in the analysis.  Recent
calculations using a bag model predict widths of
$\Gamma(\Bnull)=0.141\;\GeVcc$ and $\Gamma(\Bbr)=0.139\;\GeVcc$ for
the broad states\cite{aksel}.}

\end{center} 
\end{table}

In this paper a different approach is presented. Using information
from the photon in the decay $\bstar \rightarrow {\rm B} \gamma$, the
$\bstarjtobsb$ decays are statistically separated from the
$\bstarjtobb$ decays. This allows a model-independent measurement of
the branching ratio $\brsb$. Assuming that $\rm B^{(*)}\pi X$ decays
produce small contributions to the $\bstarj$ width, this method gives
insight into the decomposition of the $\bstarj$ into the states
allowed to decay to ${\rm B}\pi$ ($\Bnull$ and $\Btwo$) from the other
states that can only decay to $\bstar\pi$.

The $\bpi$ invariant mass spectrum is also fit in the context of HQET
expectations. This requires high statistics, extensive background
studies and the tagging of the $\bstar$ decay. Due to the complexity
of the $\bstarj$ signal and its different decay modes, several
constraints provided by HQET are imposed. From the fitting procedure,
we obtain model-dependent measurements of the mass and width of $\Bna$
as well as $\brs$.

The paper is organised as follows: the next section describes the data
sample and the event simulation. In Section~\ref{s:overview}, the
analysis method is presented. The B reconstruction is described in
Section~\ref{s:bsel}. Section~\ref{s:bstar} contains the photon
reconstruction. The pion reconstruction and the total $\bpi$ mass
spectrum are presented in Section~\ref{s:bdstar}.  The $\brsb$
measurement and results from a simultaneous fit to the $\bpipm$ mass
spectra of the samples enriched or depleted in the decay
$\bstarjtobsb$ are presented in Section~\ref{s:results}. Systematic
uncertainties are evaluated in Section~\ref{s:sys}. A discussion of
the results and conclusions are given in Section~\ref{s:sum}.
%
%
%
%
%
%
\section{Data sample and event simulation} \label{s:dsam}

The data used for this analysis were collected from $\epluseminus$
collisions at LEP during 1991--1995, with centre-of-mass energies at
and around the peak of the $\z$ resonance. The data correspond to an
integrated luminosity of about 140 ${\rm pb}^{-1}$. A detailed
description of the OPAL detector can be found
elsewhere\cite{opalsi3d,opaldet}.

Hadronic events are selected as described in \cite{pr035}, giving a
hadronic $\z$ selection efficiency of $(98.4\pm 0.4)\,\%$ and a
background of less than $0.2\,\%$. A data sample of about $\nhad$
million hadronic events is selected. Each event is divided into two
hemispheres by the plane perpendicular to the thrust axis and
containing the interaction point of the event. The thrust axis is
calculated using tracks and electromagnetic clusters not
associated with any tracks. To select events within the fiducial
acceptance of the silicon microvertex detector and the barrel
electromagnetic calorimeter, the thrust axis direction\footnote{ In
the OPAL coordinate system, the $x$ axis points towards the centre of
the LEP ring, the $y$ axis points upwards and the $z$ axis points in
the direction of the electron beam. $\theta$ and $\phi$ are the polar
and azimuthal angles, and the origin is taken to be the centre of the
detector.}  is required to satisfy $|\cos\theta_T|<0.8$.  Monte Carlo
simulated samples of inclusive hadronic $\z$ decays are used to
evaluate efficiencies and backgrounds. The JETSET 7.4 parton shower
Monte Carlo generator \cite{jetset}, with parameters tuned by OPAL
\cite{jetsetopt} and with the fragmentation function of Peterson {\it
et al.} \cite{fpeter} for heavy quarks is used to generate samples of
approximately 10 million hadronic $\z$ decays, 2 million $\ztocc$ and
5 million $\ztobb$ decays. The generated events are passed through a
program that simulates the response of the OPAL detector \cite{gopal}
before applying the same reconstruction algorithms as for data. All
generated Monte Carlo samples contain $L=1$ states for bottom and
charmed mesons, as well as vector meson partners of the
ground states. The generated production rates, masses and widths of
all resonant states are consistent with experimental measurements
when available and with theoretical predictions elsewhere (see also
Table~\ref{t:states}).
%
%
%
%
%
%
\section{Analysis overview}\label{s:overview}

The analysis is based on the reconstruction of $\bstar$ in the ${\rm
B}\gamma$ final state and a separate reconstruction of $\bstarj$ in
the $\bpipm$ final state. A direct reconstruction of $\bstarj$
decaying to $\bstar\pi$, $\bstartob$ giving ${\rm B}\gamma\pi^\pm$ in
the final state is inappropriate because of the large combinatorial
background and the insufficient detector resolution. Therefore, our
approach employs a statistical separation of $\bstarjtobsb$ from
$\bstarjtobb$ decays.

B mesons produced in $\ztobb$ events are reconstructed inclusively to
achieve high efficiency. No attempt is made to reconstruct specific B
decay channels. On the contrary, properties common to all weakly
decaying b hadrons are used for the B reconstruction. For each B
candidate, a weight $\bsprob$ is formed where $\bsprob$ represents the
probability for the B to have come from a $\bstar$. The probability $\bsprob$ is based
on the reconstruction of photon conversions and of photons detected in
the electromagnetic calorimeter. All B candidates are then combined
with charged pions to form $\bstarj$ meson candidates. Using the
weight $\bsprob$, we derive two mutually exclusive subsamples of ${\rm
B}\pi^\pm$ combinations, one enriched and the other depleted in its
$\bstar$ content. Invariant $\bpipm$ mass distributions are formed for
both samples. The shape of the non-$\bstarj$ background of the two
distributions is taken from Monte Carlo simulation and normalised to
the data in the upper sideband region and subtracted from the
corresponding data distributions. The branching ratio $\brsb$ is
obtained from the observed number of $\bstarj$ and the different
efficiencies for $\bstarjtobsb$ and $\bstarjtobb$ decays in the
$\bstar$-enriched and the $\bstar$-depleted samples.  Applying a
simultaneous fit to the $\bpi$ mass spectra of both samples several
details of the $\bstarj$ four-state composition and of the $\bstarj$
decay modes are extracted.

Whereas the $\brsb$ result obtained from counting the number of
$\bstarj$ signal entries of the samples enriched or depleted in the
decay $\bstarjtobsb$ is model-independent and does not rely on the
shape of the $\bpi$ signal, the fit to the $\bpi$ mass spectra makes
use of HQET assumptions on the composition and the decay modes of the
$\bstarj$ signal.

%
%
%
%
%
%
\section{Selection and reconstruction of B mesons}\label{s:bsel}

B mesons are reconstructed using an extended version of the method
used in earlier analyses\cite{bstar,bdstar}. Since the reconstructed B
mesons are used to form $\bstar$ and $\bstarj$ candidates, the B
reconstruction is tuned to minimise the uncertainties on the B
direction and energy, while maintaining a high reconstruction
efficiency.

\boldmath
\subsection{Tagging of $\ztobb$ events}\label{ss:btag}
\unboldmath

To achieve optimal b-tagging performance, each event is forced to a
2-jet topology using the Durham jet-finding scheme \cite{durham}. In
calculating the visible energies and momenta of the event and of
individual jets, corrections are applied to prevent double counting of
energy in the case of tracks with associated clusters
\cite{chargino172}. A b-tagging algorithm is applied to each jet using
three independent methods: lifetime tag, high $p_{\rm T}$ lepton tag and
jet-shape tag. A detailed description of the algorithm can be found in
\cite{higgs183}. The b-tagging discriminants calculated for each of
the jets in the event are combined to yield an event b likelihood
${\cal B}_{\rm event}$. For each event, ${\cal B}_{\rm event}>
\lbevcut$ is required. After this cut, the $\ztobb$ event purity is
about $\bpurity$\%. The cut on the direction of the event thrust axis,
$|\cos \theta_{\rm T}|<0.8$, as described in Section~\ref{s:dsam}, removes
roughly a quarter of all $\ztobb$ events and after the cut on ${\cal
B}_{\rm event}$, the total b event tagging efficiency \wrt\ all
produced $\ztobb$ events is about $\beffy$\%, where these numbers are
obtained from Monte Carlo simulation. At this stage, about $750\, 000$
b hadron candidates are selected.

\subsection{Reconstruction of B energy and direction}\label{ss:brec}

The primary event vertex is reconstructed using the tracks in
the event constrained to the average position and effective spread of
the $\epluseminus$ collision point. For the b hadron reconstruction,
tracks and electromagnetic calorimeter clusters with no
associated track are combined into jets using a cone algorithm
\cite{jetcone} with a cone half-angle of $0.65\rm\,rad$ and a minimum
jet energy of $5.0\rm\,GeV$ \footnote{The cone jet-finder provides the
best b hadron energy and direction resolution compared to other jet
finders studied here.}. The two most energetic jets of each event are assumed to
contain the b hadrons. For each jet we reconstruct the energy and
direction.

In each hemisphere defined by the jet axis, a weight is assigned to
each track and each cluster, where the weight corresponds to the
probability that any one track or cluster is a product of the b hadron
decay. The b hadron is reconstructed by summing the weighted momenta
of the tracks and clusters. The reconstruction algorithm is applied to
all b hadron species and is 100\% efficient. Since this analysis aims
at the reconstruction of $B_{\rm u,d}$ mesons which make up about 80\%
of the b hadron sample, b hadron candidates are referred to as B
mesons in the following. Details of the reconstruction method are
provided below.

\subsubsection{Calculation of track weights}

Two different types of weights are assigned to each track:
\begin{itemize}
  \item $\omega_{\rm vtx}$, calculated from the impact parameter
        significances of the track \wrt\  both the primary and
        secondary vertices;
  \item $\omega_{\rm NN}$, the output of a neural network based on
        kinematics and track impact parameters \wrt\  the primary
        vertex.
\end{itemize}
The calculation of $\omega_{\rm vtx}$ requires the existence of a
secondary vertex, whereas $\omega_{\rm NN}$ does not and is therefore
available for all tracks. The search for detached secondary vertices
proceeds as follows:

Each jet is searched for secondary vertices using a vertexing
algorithm similar to that described in \cite{bdstar}, making use of
the tracking information in both the $r-\phi$ and $r-z$ planes if
available. If a secondary vertex is found, the primary vertex is
re-fitted excluding the tracks assigned to the secondary vertex.
Secondary vertex candidates are accepted and called `good' secondary
vertices if they contain at least three tracks and have a decay length
greater than 0.2~mm. If there is more than one good secondary vertex
attached to a jet, the vertex with the largest number of
significant\footnote{A track is called significant if its impact
parameter significance \wrt\ the primary vertex is larger than 2.5.}
tracks is taken. If there is a tie, the secondary vertex with the
larger separation significance \wrt\ the primary vertex is taken. If a
good secondary vertex is determined, a weight is calculated for each
track in the hemisphere of the jet using the impact parameter
significance of the track \wrt\ both the primary and secondary
vertices. This weight is given by
\begin{equation}
        \omega_{\rm vtx} = \frac{R(b/\eta)}{R(b/\eta)+R(d/\sigma)}\; ,
\end{equation}
where $b$ and $\eta$ are the impact parameter and its error with
respect to the secondary vertex, and $d$ and $\sigma$ are the same
quantities with respect to the primary vertex. $R$ is a symmetric
function describing the impact parameter significance distribution
with respect to a fitted vertex.  The $\omega_{\rm vtx}$ distribution
for tracks of hemispheres with a good secondary vertex is shown in
Figure~\ref{f:trackweights}a and compared with the corresponding Monte
Carlo distribution. The weight $\omega_{\rm vtx}$ shows a weak
correlation with the momentum of the track.

For each track, the weight $\omega_{\rm NN}$ is calculated using an
artificial neural network \cite{jetnet} trained to discriminate b
hadron decay products from fragmentation tracks in a jet. The neural
network was trained using as inputs the scaled track momentum $x_p =
p/E_{\rm beam}$, the track rapidity relative to the estimated B
direction, the impact parameters of the track with respect to the
primary vertex in the $r-\phi$ and $r-z$ planes and the corresponding
errors on the impact parameters\cite{rb}. As a preliminary estimate,
the jet axis is taken as the estimated B direction. The $\omega_{\rm
NN}$ distribution is shown in Figure~\ref{f:trackweights}b. If a good
secondary vertex exists, the track weight $\omega_{\rm NN}$ is
combined with the vertex weight $\omega_{\rm vtx}$ using the
prescription
\begin{equation}
\label{weightformula}
    \omega_{\rm tr} = \frac{ \omega_{\rm NN} \cdot \omega_{\rm vtx}  }
                      {(1-\omega_{\rm NN})\cdot (1-\omega_{\rm vtx})+
                       \omega_{\rm NN} \cdot \omega_{\rm vtx}}\: .
\end{equation}
The weight $\omega_{\rm tr}$ in Equation~\ref{weightformula}
is approximately the probability that the track is a b hadron
decay product.
In the case where there is no
good secondary vertex in the jet, the total track weight $\omega_{\rm
tr}$ is simply given by $\omega_{\rm tr} = \omega_{\rm NN}$. The
combined weight $\omega_{\rm tr}$ for tracks of all hemispheres is
shown in Figure~\ref{f:trackweights}c.

\subsubsection{Calculation of cluster weights}

Similar weights are calculated for energy clusters reconstructed in
the electromagnetic and hadronic calorimeters to represent the
probability the clusters came from a B hadron decay. Weights
$\omega_{\rm ecl}$ and $\omega_{\rm hcl}$ are assigned to each
electromagnetic and hadronic cluster in the hemisphere of the B meson
based on their rapidity \wrt\ the estimated B direction. The weight is
equal to the probability, calculated using the Monte Carlo simulation
as a function of the cluster energy, that the cluster came from the
decay of a B meson. Clusters associated with a track have the
estimated energy of the track subtracted.

\subsubsection{Calculation of B direction}

The B momentum is calculated iteratively by a weighted
sum of all tracks and clusters in the hemisphere:
\begin{equation}\label{temp}
  \vec{p} = \sum^{N_{\rm track}}_{i=1} \omega_{{\rm tr},i} \cdot \vec{p}_i
          + \sum^{N_{\rm ecal}}_{i=1} \omega_{{\rm ecl},i} \cdot \vec{p}_i
          + \sum^{N_{\rm hcal}}_{i=1} \omega_{{\rm hcl},i} \cdot \vec{p}_i
\end{equation}
where $N_{\rm track}$, $N_{\rm ecal}$ and $N_{\rm hcal}$ denote the
number of tracks, electromagnetic clusters and hadronic clusters,
respectively. The rapidity calculation, for both tracks and clusters,
is initially performed relative to an estimate of the B meson
direction\footnote{ The initial input for this axis is the jet
direction calculated using tracks and unassociated electromagnetic
clusters.}. The weights are then recalculated with the rapidity
determined using the new B direction estimate.

An estimate of the B hadron direction is made for all B candidates
based on the weighted momentum sum of the tracks and clusters in the
jet. In addition, the vector from the primary vertex to the secondary
vertex yields an estimate of the B direction for those jets where a
good secondary vertex has been identified. When both estimates are
available the weighted average is taken, using the calculated
uncertainties of each direction estimate. The covariance matrices of
the primary and secondary vertices determine the error on the B flight
direction. The error on the momentum sum is estimated by removing each
term in turn from the sum in Equation~\ref{temp}, calculating the
change in the B direction caused by this omission and adding up in
quadrature the corresponding error contributions from each track and
cluster. The final estimate of the B direction is obtained by taking
the error-weighted sum of the B direction calculated with the momentum
sum method and the B direction obtained from the primary and secondary
vertex positions. The direction information in the $r-z$ plane of the
secondary vertex is only used if the vertex is built with tracks that
left at least four hits in the $z$-layers of the silicon microvertex
detector (the maximum number of these hits per track is two).

The error $\Delta \alpha$ on the weighted sum of both B direction
estimators described in the previous paragraph is a measure for the
quality of the B direction\footnote{In the case where no good
secondary vertex exists, $\Delta \alpha$ is simply given by the
uncertainty on the momentum sum.}. To improve the resolution on the B
direction, which in turn dominates the ${\rm B}\pi$ mass resolution, a
cut on $\Delta \alpha$ is imposed. Since this analysis aims at a
separation of some of the $\bstarj$ states by reconstructing different
$\bstarj$ decay channels rather than obtaining a very good $\bpi$ mass
resolution, the cut $\Delta \alpha<0.035$ is rather loose.  This cut
removes the 20\% of the B candidates with the poorest direction
resolution, mainly those with no associated good secondary vertex.

\subsubsection{Calculation of B energy}

The resolution on the
total energy of the B candidate can be significantly improved by
constraining the total centre-of-mass energy, $E_{\rm CM}$, to twice
the LEP beam energy. Assuming a two-body decay of the $\z$, we obtain:
\begin{equation}\label{beamconstraint}
        E_{\rm B} = \frac{E_{\rm CM}^2-M_{\rm recoil}^2+M_{\rm B}^2}
                         {2 E_{\rm CM}}\; ,
\end{equation}
where the mass of the b hadron is set to the B meson mass $M_{\rm
B}=5.279\;\GeVcc$ and $M_{\rm recoil}$ denotes the mass recoiling
against the B meson. The recoil mass and the recoil energy $E_{\rm
recoil}$ are calculated by summing over all tracks and
clusters\footnote{ Tracks and clusters not contained in the hemisphere
of the B meson candidate have weights $\omega_i=0$. $\omega_i$ denotes
the weight $\omega_{{\rm tr},i}$, $\omega_{{\rm ecl},i}$ and
$\omega_{{\rm hcl},i}$ for tracks, electromagnetic clusters and
hadronic clusters, respectively.} of the event weighted by
$(1-\omega_i)$ and assuming the particle masses used in the
calculation of $E_i$. To account for the amount of undetected energy
mainly due to the presence of neutrinos, the recoil mass is scaled by
the ratio of the expected energy in the recoil to the energy actually
measured:
\begin{equation}\label{recoil}
    M_{\rm recoil, new} = M_{\rm recoil, old} \cdot
          \frac{E_{\rm CM}-E_{\rm B}}{E_{\rm recoil}}
\end{equation}
where $E_{\rm B}$ is taken from Equation~\ref{beamconstraint}. The
new recoil mass value $M_{\rm recoil, new}$ obtained from
Equation~\ref{recoil} is substituted into
Equation~\ref{beamconstraint} and the calculation of $E_{\rm B}$ is
iterated. After two iterations the uncertainty on the B meson energy
is minimised. A minimum B energy of 15~GeV is required to further
improve the energy resolution.

After all these cuts, the narrower Gaussian from a two Gaussian fit
to the difference between the reconstructed and generated B meson
energy has $\sigma = 2.3\;{\rm GeV}$, and 86\% of the entries are
contained within $3\sigma$.
The distribution of the difference between the reconstructed and
generated $\phi$ angle of simulated B mesons can be described by a
similar fit. The standard deviation of the narrower Gaussian is
14.2 mrad and 88\% of the entries lie within $3\sigma$. The
corresponding quantities describing the $\theta$ resolution are
$\sigma = 15.0$ mrad and 89\%, respectively.

The complete B meson selection applied to the full data sample results
in $574\,288$ tagged jets with a b purity of about 96\%, as estimated
from Monte Carlo. About 75\% of the selected jets contain a good
secondary vertex.
%
%
%
%
%
%
\boldmath
\section{The decay $\bstartob$}\label{s:bstar}
\unboldmath

The photon produced in the decay $\bstartob$ has an energy of about
46~MeV in the rest frame of the $\bstar$.  The mean energy of the
photon in the laboratory frame is approximately 350~MeV, with a
maximum energy below 800~MeV. Due to the kinematics of the process,
these photons are produced predominantly in the core of the jet. The
high particle density in this region gives rise to a high background
level when identifying the photon. Since a high $\bstar$
reconstruction efficiency is crucial for this analysis, photons are
reconstructed in two ways: from energy deposits in the electromagnetic
calorimeter and from converted photons in the tracking volume. The
conversion probability within the OPAL tracking system for photons
coming from the decay $\bstartob$ is approximately~8\%.

\subsection{Reconstruction of photon conversions}\label{ss:conv}

The reconstruction of converted photons is optimised for the low
energy region. The selection algorithm is partially based on
quantities that have been used in earlier analyses \cite{conversions}
but tuned to obtain high efficiency rather than very good angular and
momentum resolution. Given the low energy carried by these photons, we
ignore calorimetry information and only use tracking information for
the reconstruction of converted photons.

Tracks with a total momentum $p$ below 1.0~$\GeVc$, opposite charge
assignment
and a measured ${\rm d}E/{\rm d}x$ within three standard deviations of
the expected value for electrons are combined into pairs. For each
pair, the track with the greater scalar momentum is required to have a
transverse momentum $p_t > 50\;\MeVc$ \wrt\ the beam axis and at least
20 hits in the central jet chamber. For the track with lower momentum,
a minimum $p_t$ of $20\;\MeVc$ is required. The asymmetric selection
cuts for the two tracks in a pair guarantee at least one well measured
track and reflect the fact that the electron and the positron of a
converted photon tend to have different momenta in the laboratory
frame. To suppress random track combinations, the distance of closest
approach between the two tracks of a pair in the $r-\phi$ plane has to
be smaller than 1.0 cm with an opening angle between the tracks at
their point of closest approach smaller than 1.0 rad.

In order to make optimal use of all the available information, the
following measured quantities for each conversion track pair candidate
are fed into a neural network:
\begin{itemize}
\item the distance of closest approach between the two tracks in the
$r-\phi$ plane;
\item the radial distance \wrt\  the $z$ axis of the first and last
measured hits in the inner tracking chambers for each track;
\item the radial distance \wrt\  the $z$ axis of the common
vertex\footnote{The $z$ position of this vertex is fitted
independently and the reconstructed photon vector is constrained to
the $z$ coordinate of the primary vertex to improve the accuracy of
the $\theta$ determination.} of both tracks obtained from a fit in the
$r-\phi$ plane;
\item the impact parameter \wrt\  the primary vertex in the $r-\phi$
plane of the reconstructed photon;
\item the invariant mass of the track pair assuming both tracks to be
electrons;
\item the transverse momentum relative to the $z$ axis of the lower
momentum track.
\end{itemize}
All conversion candidates with a neural network output greater than 0.7
and a photon energy below 1.5~GeV are called `good' conversion
candidates for a given B meson candidate if the opening angle between
the reconstructed B momentum vector and the reconstructed photon
momentum vector is smaller than $90^\circ$. At this stage, an average
of 0.82 good conversion candidates is selected per B candidate in
both data and Monte Carlo. The candidate multiplicity distributions are shown
in Figure~\ref{f:ncand}a. The total efficiency to detect photons from
the decay $\bstartob$ with the conversion algorithm is estimated from
simulation to be $(2.70\pm0.01_{\rm stat})\%$. The efficiency is
rather independent of the photon energy from 1.0~GeV down to 200~MeV
where it rapidly drops to zero due to track selection
requirements. The amount of fake conversions in the selected sample is
estimated from Monte Carlo simulation to be $(11.75\pm0.04_{\rm
stat})\%$.

Fits to the difference between the reconstructed and generated photon
energy in Monte Carlo are made using the sum of two Gaussians, both
constrained to the same mean value. The narrower Gaussian has a
standard deviation of 5~MeV at an energy of 200~MeV, and rises to
13~MeV at an energy of 750~MeV, and about 70\% of the entries are
contained within $3\sigma$. Similar fits to the $\phi$ and $\theta$
resolutions give values of 3.4 mrad (70\%) and 5.4 mrad (61\%),
respectively.

\subsection{Reconstruction of photons in the electromagnetic calorimeter}\label{ss:ecal}

Photons are also detected as showers in the barrel region of the
electromagnetic calorimeter. The location and energy of these showers
are obtained from a fit to the energy deposits in the individual lead
glass blocks not associated with any track. The whole
reconstruction method has been shown to work in the dense environment
of hadronic jets down to photon energies as low as 150~MeV.  The
details of the reconstruction are given in \cite{gascon}.

Showers in the electromagnetic calorimeter are accepted as photon
candidates if they have an energy in the range 200~MeV to 850~MeV and
a photon probability $P_\gamma > 0.20$, where $P_\gamma$ is the output
of a simplified neural network\cite{gascon}. If the opening angle
between such a shower and a reconstructed B candidate is less than
$90^\circ$, this shower is considered a `good' photon candidate for
the corresponding B candidate. On average, there are 4.59 (4.38) good
calorimeter photon candidates per B candidate selected in the data
(Monte Carlo) sample. To correct for the observed discrepancy, the
Monte Carlo is reweighted to the data distribution shown in
Figure~\ref{f:ncand}b. The efficiency to detect a photon from the
decay $\bstartob$ in the electromagnetic calorimeter is estimated to
be $(14.52\pm 0.03_{\rm stat})\%$ using Monte Carlo simulated events.
The fraction of fake photons arising from tracks and neutral
hadrons in the sample ranges from 32\% at a photon energy of 850~MeV
up to 43\% at a photon energy of 200~MeV. If compared with the
selected conversion sample, the selection of $\bstartob$ photons in
the electromagnetic calorimeter has a much higher efficiency but lower
purity.

As with the converted photons, the energy resolution has been
determined from Monte Carlo simulation using a double Gaussian fit.
The narrower Gaussian has a width of 20 MeV at a photon energy of 250
MeV and increases up to 86 MeV at an energy of 800 MeV, and about 75\%
of the entries are contained within $3\sigma$. Similar fits to the
$\phi$ and $\theta$ resolutions give values of 3.6 mrad (65\%) and 3.6
mrad (72\%), respectively. In contrast to the conversion sample,
photons reconstructed in the electromagnetic calorimeter have much
higher energy uncertainties, but a better $\theta$ resolution.

\boldmath
\subsection{Reconstruction of $\bstar \rightarrow {\rm B}
                           \gamma$ decays}\label{ss:bsrec}
\unboldmath

Each reconstructed B meson candidate is combined with all good
conversion and calorimeter candidates to reconstruct $\bstar$ 
candidates. The invariant mass of a ${\rm B}\gamma$ combination is
defined as
\begin{equation}\label{bstarmass}
M_{{\rm B}\gamma} = \sqrt{M_{\rm B}^2+2E_{\rm B}E_{\gamma}
                       -2p_{\rm B}2p_{\gamma} \cos \alpha } \quad,
\end{equation}
where $M_{\rm B}$ is 5.279~$\GeVcc$ and $\alpha$ is the measured angle
between the B meson and the photon candidate. The mass difference
$\dmbstar = M_{{\rm B}\gamma}-M_{\rm B}$ between the $\bstar$ candidate
and the B is calculated by simply subtracting the nominal B mass of
$M_{\rm B}=5.279\;\GeVcc$ from $M_{{\rm B}\gamma}$.

The mass difference distributions of the conversion sample observed in
the data and the corresponding Monte Carlo background are shown in
Figure~\ref{f:bsconv}a. The background is normalised to the data in
the sideband region $0.09\; \GeVcc < \dmbstar < 0.20\; \GeVcc$. The
background subtracted signal of Figure~\ref{f:bsconv}b is fitted
\label{p:bsconvfit} to the sum of two Gaussians fixed to the same
mean, where one of the Gaussians is allowed to have asymmetric
width. The observed asymmetry of the mass resolution of the conversion
sample is well simulated in the Monte Carlo and is due to the very
loose track requirements of the lower momentum track of the conversion
pair. A mass difference of $\dmbstar = (45.87 \pm 0.25_{\rm stat})\;
\MeVcc$ is obtained from the fit to the data, where the error is
statistical only. This result agrees well with the current world
average value of $(45.78 \pm 0.35)\; \MeVcc$ \cite{pdg}.

The $\dmbstar$ distribution using calorimeter photons is shown in
Figure~\ref{f:bsecal}a. The background is taken from Monte Carlo
simulation and normalised to the data in the sideband region $0.10\;
\GeVcc < \dmbstar < 0.20\; \GeVcc$. The same fit function
\label{p:bsecalfit} as for the photon conversion sample is used to
obtain the mass difference $\dmbstar$ from the background subtracted
signal distribution in Figure~\ref{f:bsecal}b. A value of $(47.30 \pm
0.61_{\rm stat})\; \MeVcc$ is obtained from this fit, which is
consistent within two sigma with the result from the fit to the
conversion signal.

For the $\bstar$ sample reconstructed with photon conversions, the
mass resolution is dominated by the uncertainty on the reconstructed B
direction. For the calorimeter photon sample, the $\bstar$ mass
resolution suffers in addition from the energy resolution of the
calorimeter. Due to the high background of fake photons and the
moderate energy resolution at low photon energies, the
signal-to-background ratio is rather poor for calorimeter photons.
Therefore, uncertainties in the $\bstar$ reconstruction using these
photons are dominated by systematic errors on the background shape and
energy calibration. All systematic uncertainties arising from the
$\bstar$ reconstruction will be discussed in Section~\ref{s:sys}.

\boldmath
\subsection{The $\bstar$ probability $\bsprob$}\label{ss:enrdep}
\unboldmath

To select samples enhanced and depleted in $\bstar$ mesons, a $\bstar$
probability is assigned to each B candidate. This probability combines
information from both conversion and calorimeter photon candidates and
represents the probability that a B candidate is the true daughter of
a $\bstar$ meson. Only the best conversion and best calorimeter
candidate assigned to any one B candidate are considered in the
calculation of this probability, where the best candidate is defined
as that which gives $\Delta M=M_{\rm B\gamma} -M_{\rm B}$ closest to
the world average of $45.78\;\MeVcc$ \cite{pdg}.

This weight is constructed by parametrising the purity of the mass
difference distribution in several variables in Monte Carlo
simulation. For calorimeter photon candidates, this parametrisation is
performed as a function of the photon probability, $P_\gamma$ (see
Section~\ref{ss:ecal}), and the total number of good calorimeter
photon candidates found per B candidate. For each B candidate, a single
weight is calculated by taking the simple mean of the weight resulting
from each of the above parametrisations.

Similarly, for conversion photon candidates, the parametrisation is
performed in $\Delta M$ as a function of the total number of
conversion candidates, and a weight is extracted as for the
calorimeter candidates. The two weights obtained from conversion and
calorimeter photons are combined by taking their mean.

The resulting weight $\bsprob$ is shown in Figure~\ref{f:bsprob}a for
Monte Carlo and data, and the contributions from jets containing a
$\bstar$ and jets containing no $\bstar$ as seen in the simulation are
shown. The primary features of the $\bsprob$ distribution are:
\begin{itemize}
\item a peak at $\bsprob = 0.625$, corresponding to B candidates with
no associated good conversion or calorimeter photon candidate;

\item a peak at $\bsprob = 0.632$, containing B candidates with no
good conversion candidate and a best calorimeter candidate having a
${\rm B}\gamma$ mass far away from the nominal $\bstar$ mass;
  
\item a peak around $\bsprob = 0.656$, containing B candidates with
the best calorimeter candidate close to the nominal $\bstar$ mass;

\item a peak at $\bsprob = 0.715$, containing B candidates with the
best calorimeter candidate being close to the nominal $\bstar$ mass and
having a high photon probability $P_\gamma$;
  
\item the tail towards high $\bstar$ probabilities is made up by best
conversion candidates very close to the nominal $\bstar$ mass.
\end{itemize}

The assignment of the specific photon candidate samples to the peaks
and to the tail of the $\bsprob$ distribution is based on Monte Carlo
information.  Figure~\ref{f:bsprob}b shows the ratio
$\varepsilon(\bstar)/\varepsilon(\rm B)$ versus $\bsprob$.
$\varepsilon(\bstar)$ refers to the efficiency to select a B meson
from a true $\bstartob$ decay and $\varepsilon(\rm B)$ is the
efficiency to select a B meson which has not come from a $\bstar$.  In
general, the Monte Carlo simulation of $\bsprob$ describes the data
adequately. A comparison of the Monte Carlo and data distributions
yields a $\chi^2$ per degree of freedom of two. A cut on $\bsprob$
allows the production of samples of B candidates with different
$\bstar$ fractions. Further details and systematic studies concerning
$\bsprob$ are given in Section~\ref{s:sys}.
%
%
%
%
%
%
%
%
\boldmath
\section{Reconstruction of orbitally-excited B mesons}\label{s:bdstar}
\unboldmath

All $\bstarj$ candidates, even those expected to decay into
$\bstar\pi^\pm$, are reconstructed using the measured four-momenta of
the B meson and the pion. B candidates are selected and reconstructed
as described in Section~\ref{s:bsel} and combined with charged pion
candidates. Pions produced in the decay of a $\bstarj$ will be
referred to as `signal pions'. Since the $\bstarj$ decays strongly,
signal pions are expected to be associated to the primary event vertex
rather than to a possible secondary vertex. In comparison with other
pions created in the fragmentation process, signal pions are expected
to have a large longitudinal momentum $p_l$ with respect to the jet
axis. These are the basic characteristics used to separate signal
pions from B decay products and from fragmentation tracks. A
significant number of non-resonant fragmentation pions are expected to
be produced near a B meson. The kinematics of these pions is similar
to the signal pions, giving rise to a combinatorial background in the
invariant mass of ${\rm B}\pi$ candidates. Tracks from B decay also
contribute to the background due to the inability to unambiguously
associate all B decay tracks with the secondary vertex.

\subsection{Pion selection}\label{ss:pionselection}

The signal pion selection for this analysis makes use of techniques
used in \cite{bdstar} and \cite{bdsalephexcl}. All tracks that
are well measured according to a standard track selection
\cite{tracksel} are considered as possible signal pion candidates if
they belong to the same jet as the B candidate. Additionally, the
following selection cuts are applied in the given order:
\begin{itemize}
\item The measured ionisation energy loss ${\rm d}E/{\rm d}x$ has to
be consistent with the expected value for pions within 2.6 standard
deviations, if ${\rm d}E/{\rm d}x$ information is available for this
track.
\item To suppress B decay tracks, the track weight $\omega_{\rm NN}$
as described in Section~\ref{ss:brec} has to be smaller than 0.9.
\item The B decay track rejection is improved by the requirement
$\omega_{\rm NN2}<0.7$, where $\omega_{\rm NN2}$ is a neural network
output defined for jets containing a secondary vertex. The inputs for
$\omega_{\rm NN2}$ are similar to the inputs for $\omega_{\rm NN}$,
but also the impact parameter significances in the $x-y$ and the $z$
plane \wrt\  the secondary vertex are used.
\item From all tracks that pass the previous selection criteria, only
the one with the highest longitudinal momentum \wrt\  the jet axis,
$p_l^{\rm max}$, is kept for each B candidate,

\item A reduction of B decay track background in the $p_l^{\rm max}$
sample is obtained by the requirements $\omega_{\rm NN}<0.80$ and
$\omega_{\rm NN2}<0.50$ \footnote{If no secondary vertex is present in
the jet, the cut $\omega_{\rm NN}<0.50$ instead of $\omega_{\rm
NN2}<0.50$ is applied.}.
\item Fragmentation tracks in the $p_l^{\rm max}$ sample are removed
with the requirement $\omega_{\rm NN}>0.20$. Since $\omega_{\rm NN}$
is designed to achieve optimal separation of b hadron decay tracks
from fragmentation tracks using impact parameter information {\it and}
kinematics, the Monte Carlo indicates a fairly flat $\omega_{\rm NN}$
distribution for signal pions. On the contrary, fragmentation tracks
peak at zero.

\item A momentum of $p>1.0\;\GeVc$ is required for signal pion
candidates. In the simulation, the momentum distribution of signal
pions has a mean value of $2.9\;\GeVc$ with an RMS of $1.3\;\GeVc$
before the cut is applied. The $\bstarj$ mass spectrum for single pion
transitions is not influenced by the momentum requirement.
\end{itemize}

\boldmath
\subsection{$\bpipm$ mass spectrum} \label{ss:bpimspectrum}
\unboldmath

The signal pion candidate passing the selection cuts described in
Section~\ref{ss:pionselection} is combined with the corresponding B
candidate to form a $\bstarj$ candidate. The invariant mass is
calculated using Equation~\ref{bstarmass} as for the $\bstar$ mass,
where the photon is replaced by a pion and the appropriate pion mass
term is added. The cuts of the signal pion selection have been chosen
to obtain an acceptable signal-to-background ratio at high signal
efficiency. The order of the non-commuting selection requirements
using $\omega_{\rm NN}$, $\omega_{\rm NN2}$ and $p_l^{\rm max}$ aims
to maximise the difference between the shape of the signal and
background contributions to the $M_{\rm B\pi}$ distribution.

Due to the intrinsic widths of the $\bstarj$ states and the limited
detector resolution, only a single peak is seen in the $M_{{\rm
B}\pi}$ spectrum of Figure~\ref{f:bdsall}a on top of the combinatorial
background. According to the simulation, the $M_{\rm B\pi}$ resolution
can be described by the sum of a narrow Gaussian and an asymmetric
Gaussian, both constrained to the same mean value (see
Figure~\ref{f:bpireso}). The mass resolution depends linearly on
$M_{\rm B\pi}$. In the $\bstarj$ signal region around $5.7\;\GeVcc$,
the standard deviation of the narrow Gaussian is $\sigma=33\;\MeVcc$,
and 85\% of the resolution function entries are contained within
$3\sigma$. The reliability of the simulated B meson energy and
direction resolution which dominate the $\rm B\pi$ mass resolution is
proven by a well simulated shape and peak position of the $\bstar$
signal using the conversion photon sample (see Section \ref{s:bstar}).

The Monte Carlo combinatorial background is checked against data using
different test samples strongly enhanced in each of the following
physics background sources: 1) Fake $\bstarj$ candidates from light
and charm quark events; 2) Fake $\bstarj$ arising from true b hadrons
combined with a pion from the weak decay of the b hadron itself;
3) Fake $\bstarj$ formed by combining true b hadrons with
fragmentation tracks which have not come from a $\bstarj$ resonance.
The simulation indicates that each test sample is strongly enhanced in
the background source under study and that the $\bstarj$ signal is
suppressed by about a factor of eight compared to the original
$\bstarj$ signal selection. The $\rm B\pi$ mass distributions of the
background samples in data are compared with the corresponding Monte
Carlo mass distributions (see also Section~\ref{ss:syspion}). In the
case of a significant deviation, the simulated background is
reweighted to the data. The Monte Carlo background distribution so
obtained is fitted \label{p:bdsallfit} using a threshold function of
the form
\begin{equation}
C_1 \cdot \sqrt{x-(m_{\rm B}+m_\pi)} \cdot \left( \Phi
\left( \frac{x-C_2}{C_3}\right) \right)^{C_4} \quad ,
\end{equation}
where $\Phi$ is the Landau density
function\footnote{$\Phi(\lambda) =\frac{1}{2\pi i}
\int_{c-i\infty}^{c+i\infty} e^{\lambda s + \ln s} ds$ }. This
background function gives a good empirical fit with only four free
parameters $C_i$. The fitted Monte Carlo background is normalised in
the sideband region $6.10\;\GeVcc < M_{{\rm B}\pi} < 7.10\;\GeVcc$ and
subtracted from the data distribution. The obtained $\bstarj$ signal
is shown in Figure~\ref{f:bdsall}b.

The reconstruction efficiency for $\bstarj$ depends on the
reconstructed mass $M_{\rm B\pi}$. Monte Carlo studies indicate that
the efficiency stays constant at high $M_{\rm B\pi}$ values down to
$M_{\rm B\pi}=5.7\;\GeVcc$. Below $5.7\;\GeVcc$, the reconstruction
efficiency becomes smaller as $M_{\rm B\pi}$ decreases, mainly due to
the $p_l^{\rm max}$ requirement.  At the $\rm B\pi$ mass threshold,
the signal efficiency is close to zero.  The $\bstarj$ distribution of
Figure~\ref{f:bdsall}b is corrected for efficiency and the resulting
signal is shown in Figure~\ref{f:bdsall}c.

The mean mass, shape and yield of the observed $\bstarj$ signal is
roughly in agreement with other measurements \cite{bdstar,
bdstardelphi, bdsalephincl, bdsalephexcl}. The structure of the
$\bpipm$ mass spectrum is too broad to stem from a single resonance
and leaves room for interpretation. The peak is expected to contain
two broad and two narrow $\bstarj$ states, and part of the true mass
spectrum is shifted to lower mass values by $46\;\MeVcc$ due to the
omission of the photon in the reconstruction of $\bstarj\rightarrow
\bstar \pi$ decays. The peak may also include a small fraction of
$\bstarjs$ due to the misidentification of kaons as pions. In
addition, the peak may contain contributions from $\bstarj \rightarrow
{\rm B}^{(*)}\pi X$ giving rise to satellite peaks in the region
$5.4\;\GeVcc < M_{\rm B\pi} < 5.6\;\GeVcc$, since $X$ is not included
in the invariant mass calculation. If broad $\bstarj$ states have
masses close to the $\rm B\pi$ threshold, they have an asymmetric
signal shape due to phase-space suppression. Radially-excited B mesons
(2S) in the decay channels ${\rm B}^{(*)\prime} \rightarrow {\rm
B}^{(*)}\pi$ and ${\rm B}^{(*)\prime} \rightarrow {\rm B}^{(*)}\pi\pi$
may contribute, although the production rate of ${\rm B}^{(*)\prime}$
is assumed to be small compared to the $\bstarj$ production rate
according to \cite{bdstarl3} and \cite{bspsup}. Since there are
several ambiguities, e.g. due to $\bstarj \rightarrow {\rm
B}^{(*)}\pi\pi$, ${\rm B}^{(*)\prime}$ decays and uncertainties in the
combinatorial background, further details of the signal can only be
obtained by making additional, model-dependent assumptions.
%
%
%
%
%
%
\boldmath
\section{$\bstarj$ transitions to $\bstar$ and to B}\label{s:results}
\unboldmath

In this section, information from the reconstructed $\bpipm$ mass is
combined with the weight $\bsprob$. The total $\bstarj$ sample is
divided into two samples, one enriched and one depleted in the decay
$\bstarj\rightarrow\bstar\pi(X)$, by applying a cut on $\bsprob$, as
indicated in Figure~\ref{f:bsprob}a. These two $\bstarj$ samples allow
a model-independent measurement of the ratio $\brsb$, where no
distinction between decays to $\bstar\pi$ and $\bstar\pi X$ is
possible. Also a fit with $\brs$ as an additional free fit parameter
is performed. In this fit, only $\bstarj\to{\rm B}^{(*)}\pi$ and
$\bstarj\to{\rm B}^{(*)}\pi\pi$ decays are considered. The fit result
of the branching ratio $\brs$ does {\em not} include the $\bstarpipi$
final state as the sensitivity of the fit is negligible for di-pion
transitions of the $\bstarj$.

\boldmath
\subsection{Model-independent measurement of $\brsb$}
\label{ss:modindbrs}
\unboldmath

The branching ratio $\brsb$ is obtained by counting the number of
signal entries of the $\bstarj$ samples enriched or depleted in the
decay $\bstarjtobsb$. The cut value on $\bsprob$ is chosen to minimise
the uncertainty of the measurement. The statistical error on $\brsb$ is
minimal if both subsamples are of the same size. Systematic
uncertainties in the $\bpipm$ background have minimal impact on
$\brsb$ if the signal-to-background ratio is the same for the $\bpipm$
mass distributions of the $\bstar$-enriched and the $\bstar$-depleted
samples. The optimal cut on $\bsprob$ is $0.648$, fulfilling the
minimal systematic error requirement and coming as close as possible
to the minimum statistical error requirement (see
Figure~\ref{f:bsprob}).

The $\bstar$ enrichment and depletion method can be evaluated by the
different selection efficiencies for the transitions $\bstarjtobs$ and
$\bstarjtob$ in the $\bstar$-enriched and $\bstar$-depleted samples.
With the definitions
\begin{itemize}
     \item $\varepsilon_{\rm _E}^*\:$: \quad
           $\bstarj\rightarrow\bstar\pi$ efficiency of
           $\bstar$-enriched sample;

     \item $\varepsilon_{\rm _D}^*\:$: \quad
           $\bstarj\rightarrow\bstar\pi$ efficiency of
           $\bstar$-depleted sample;

     \item $\varepsilon_{\rm _E}\:$: \quad $\bstarj\rightarrow{\rm
           B}\pi$ efficiency of $\bstar$-enriched sample;

     \item $\varepsilon_{\rm _D}\:$: \quad $\bstarj\rightarrow{\rm
           B}\pi$ efficiency of $\bstar$-depleted sample;

     \item efficiency ${\rm ratios}\:$: $e_0=\varepsilon_{\rm _D}/
           \varepsilon_{\rm _E}^*\,$; $\: e=\varepsilon_{\rm _E}/
           \varepsilon_{\rm _D}\,$; $\: e^*=\varepsilon_{\rm _D}^*/
           \varepsilon_{\rm _E}^*\;$,
\end{itemize}
we calculate from Monte Carlo the efficiency values presented in
Table~\ref{t:effy}. The numbers reflect the cut on $\bsprob$ and thus
the quality of the $\bstar$ enrichment versus the $\bstar$ depletion.
Only the efficiency ratios given in the right column of
Table~\ref{t:effy}, not the absolute efficiencies, are needed for the
determination of $\brsb$.
\begin{table}[bth]
\begin{center}
\begin{tabular}{|l|l||l|l|}

\hline
 \multicolumn{2}{|c||}{efficiency} & \multicolumn{2}{c|}{efficiency ratio} \\
\hline
$\quad \varepsilon_{\rm _E}^*\quad $ & $\quad \ese \pm \dese \quad$ &
                   $\quad e_0\quad $ & $\quad \effnull \pm \deffnull \quad$\\
$\quad \varepsilon_{\rm _D}^*\quad $ & $\quad \esd \pm \desd \quad$ &
                     $\quad e\quad $ & $\quad \eff \pm \deff \quad$        \\
$\quad \varepsilon_{\rm _E}\quad $   & $\quad \ee \pm \dee \quad$ &
                   $\quad e^*\quad $ & $\quad \effstar \pm \deffstar \quad$\\
$\quad \varepsilon_{\rm _D}\quad $   & $\quad \ed \pm \ded \quad$ &
                            $\quad $ &                                     \\
\hline
\end{tabular}
\caption{\label{t:effy} Efficiencies for the reconstruction of
$\bstarj$ decaying to $\bstar\pi$ and to ${\rm B}\pi$. The numbers are
calculated \wrt\  the total number of $\bstarj$ passing the B selection
in the Monte Carlo. Therefore, the numbers reflect the effect of the
cut on $\bsprob$ and the charged pion selection. Also a factor of 2/3
assuming isospin symmetry to account for decays of $\bstarj$ via
neutral pions is included in each of the efficiency values. The errors
are statistical only.}
\end{center} 
\end{table}
For the $\brsb$ measurement, the invariant $\bpipm$ mass distributions
of the $\bstar$-enriched and the $\bstar$-depleted sample are used.
Both mass distributions are independent subsamples of the distribution
shown in Figure~\ref{f:bdsall}, but contain different compositions of
$\bstarjtobsb$ and $\bstarjtobb$ decays. Figures~\ref{f:bsenr} and
\ref{f:bsdep} show the $\bpipm$ mass distributions for the
$\bstar$-enriched and the $\bstar$-depleted sample, respectively. The
Monte Carlo background distributions of both samples are corrected
using a procedure explained in Sections~\ref{s:bdstar} and \ref{s:sys}
and the same fit as for the background of the total $\bstarj$ sample
(see Section~\ref{s:bdstar}) is performed. The fitted background
functions are normalised in the sideband region $6.10\;\GeVcc <
M_{{\rm B}\pi} < 7.10\;\GeVcc$ and subtracted from the corresponding
data distributions. From the resulting signal peaks,
$\brsb$ is obtained. Assuming that $\bstarj$ decay to $\bstar$ or $\rm
B$ only and with the efficiency ratios defined as above, we derive the
following formula:
\begin{eqnarray}
\label{formula:brs}
     \brsb & = &
     \frac{{\rm BR}({\rm B}^{*}_{J} \rightarrow {\rm B}^{*} \pi (X))}
     {{\rm BR}({\rm B}^{*}_{J} \rightarrow {\rm B}^{*} \pi (X)) + 
     {\rm BR}({\rm B}^{*}_{J} \rightarrow {\rm B} \pi (X))} \nonumber \\
      & = &
     e_0 \cdot \frac { N_{\rm E} - e \cdot N_{\rm D}}
      { (e_0 - e^*) \cdot N_{\rm E} + 
      (1 - e \cdot e_0) \cdot N_{\rm D} } \quad ,
\end{eqnarray}
where $N_{\rm E}$ ($N_{\rm D}$) denotes the number of $\bstarj$ signal
entries of the $\bstar$-enriched ($\bstar$-depleted) sample. In the
data, $N_{\rm E}=(\None \pm \dNone_{\rm stat})$ and $N_{\rm D} =
(\Ntwo \pm \dNtwo_{\rm stat})$ $\bstarj$ candidates are observed in
the $M_{\rm B\pi}$ signal window $(5.3-6.1)\;\GeVcc$. Using the
numbers for the efficiency ratios $e_0$, $e$ and $e^*$ presented in
Table~\ref{t:effy}, we calculate $\brsb = \brval$.

The statistical errors on $N_{\rm E}$ and $N_{\rm D}$ result in a
total error on the branching ratio of $\pm \brstat$. Besides this
error, statistical uncertainties due to the sideband normalisation
have been taken into account. Since the samples are mutually
exclusive, the statistical errors of the sideband normalisation of
both samples are independent. The contributions of the
$\bstar$-enriched and the $\bstar$-depleted sample sideband
normalisation to the statistical error on $\brsb$ are $^{+0.17}_{-0.18}$
and $\pm 0.15$, respectively. Adding all quoted errors in quadrature,
the branching ratio of orbitally-excited B mesons decaying into
$\bstar$ is measured to be \vspace{0.2cm}
\[
        \brsb = \brval^{\,\brstatplus}_{\,\brstatminus}\; ,
\vspace{0.2cm}
\]
where the error is statistical only. This branching ratio includes all
decays of the type $\bstarj \rightarrow \rm B^{(*)}\pi X$, as no cut
against additional $\bstarj$ decay products is applied. Consequently,
the notation $\brsb$ is chosen. Systematic uncertainties, especially
of the efficiency ratios and the combinatorial $\rm B\pi$ background
are discussed in Section~\ref{s:sys}.

We further investigate the composition of the $\bstarj$ sample by
splitting the sample into $\bstarjtobsb$ and $\bstarjtobb$ components.
By subtracting from the ${\rm B}\pi$ mass distribution of the
$\bstar$-enriched sample the corresponding distribution of the
$\bstar$-depleted sample multiplied by a scale factor, a $\bpipm$ mass
distribution containing $\bstarjtobsb$ transitions only is obtained.
The scale factor is the ratio of the $\bstarjtob$ efficiencies of both
samples, $e=\varepsilon_{\rm _E}/\varepsilon_{\rm _D}$. In a similar
way, a mass distribution with $\bstarjtobsb$ decays subtracted off is
obtained. The corresponding efficiency-corrected $\bpipm$ mass
distributions for pure $\bstarjtobsb$ and pure $\bstarjtobb$
transitions are shown in Figure~\ref{f:pure}. The number of signal
entries in the $\bpipm$ mass distributions of Figures~\ref{f:pure}a
and~\ref{f:pure}b depends on the ratio $\brsb$ as well as on the
efficiency ratios defined in Table~\ref{t:effy}.

A significant excess of entries is seen in the pure $\bstarjtobsb$
distribution at masses around $5.7\; \GeVcc$ with tails down to
$5.5\;\GeVcc$ and up to $6.0\;\GeVcc$. The narrow peak in the
$\bstarjtobsb$ distribution is most likely due to $\rm \Bna
\rightarrow B^*\pi^\pm$ and $\Btwo\rightarrow {\rm B}^*\pi^\pm$
decays. To obtain the true mass values of the $\bstar\pi$ states, the
entries have to be shifted to higher masses by $46\;\MeVcc$.

In the pure $\bstarjtobb$ mass distribution, a small excess is
observed in the region up to $5.85\;\GeVcc$.  This excess can be
assigned to the decays $\Btwo\rightarrow {\rm B}\pi^\pm$ and
$\Bnull\rightarrow {\rm B}\pi^\pm$. Since the statistical significance
of the excess in the $\bstarjtobb$ mass distribution is small, no
further conclusion is drawn from Figure~\ref{f:pure}b.

\boldmath
\subsection{Simultaneous fit to the $\bpipm$ mass spectra}\label{ss:simulfit}
\unboldmath

A simultaneous fit is performed to the background subtracted and
efficiency corrected $\rm B\pi$ mass spectra shown in
Figures~\ref{f:bsenr}c and~\ref{f:bsdep}c. Several assumptions are
made on the nature of the observed signal to reduce the number of
free fit parameters:

\begin{itemize}

 \item The signal excess stems from $\bstarj$ decays only.
 Contributions from $\bstarjs$ decays, ${\rm B}^{(*)\prime}
 \rightarrow {\rm B}^{(*)}\pi$ and ${\rm B}^{(*)\prime} \rightarrow
 {\rm B}^{(*)}\pi\pi$ do not exceed a few percent \cite{bdstar,
 bstarprime, bdstarl3} and are therefore not implemented in the fit
 but considered as sources of systematic uncertainties. Any other
 excited state eventually contributing to the signal peak is ignored
 since there is no experimental evidence for such states and
 theoretical predictions give negligible production rates.

 \item The heavy quark limit $m_Q \rightarrow \infty$ holds to
 describe the four $\bstarj$ states. Therefore, according to
 spin-parity conservation, one expects five different mass peaks from
 single pion transitions as listed in the first paragraph of
 Section~\ref{s:intro} and shown in Figure~\ref{f:transition}.
 Furthermore, the physical $\Bone$ states are $\Bbr$ and $\Bna$ and
 thus no mixing occurs.
 
 \item Partners of the same doublet are assumed to have similar
 properties. The constraints on masses, widths and production rates
 used in the fit are presented in Table~\ref{t:fitconstraints}. The
 mass splitting between the narrow states can be calculated using the
 corresponding mass splitting of the ${\rm D}^*_J$ which has been
 measured\cite{pdg}. The mass splitting between the broad states is
 expected to be of about the same size. Also the order of magnitude of
 the widths of the narrow and broad $\bstarj$ states can be estimated
 from experimental ${\rm D}^*_J$ results \cite{pdg,broaddstarj}.

\begin{table}[tb]
\begin{center}
\begin{tabular}{|l|c||c|c|c|c|}

\hline
         &       & \multicolumn{3}{c|}{fit constraints}& \raisebox{-0.25ex}{allowed}  \\ \cline{3-5}
\raisebox{1.5ex}[-1.5ex]{state} & \raisebox{1.5ex}[-1.5ex]{$J_j^P$} & 
 production rate & mass  & width & \raisebox{0.25ex}{decay modes} \\
\hline
$\Bnull$ &  $0_{1/2}^{+}$ & $\rm f(b\rightarrow\Bnull)$& free                   &
            free                                       & $\rm B\pi$, $\bstarpipi$, $\bpipi$   \\
$\rm B_1(\frac{1}{2})$  &  $1_{1/2}^{+}$ & $\rm f(b\rightarrow\Bnull)$& $M(\Bnull)+20\;\MeVcc$ &
            $1.25\cdot\Gamma(\Bnull)$                  & $\bstar\pi$, $\bstarpipi$, $\bpipi$   \\
$\rm B_1(\frac{3}{2})$  &  $1_{3/2}^{+}$ & $\rm \frac{3}{2}\cdot f(b\rightarrow\Bnull)$  & free  &
            free                                       & $\bstar\pi$, $\bstarpipi$, $\bpipi$   \\
$\Btwo$  &  $2_{3/2}^{+}$ & $\rm \frac{3}{2}\cdot f(b\rightarrow\Bnull)$& $\!\!M(\rm B_1(\frac{3}{2}))+12\;\MeVcc\!\!$   &
            $1.00\cdot\Gamma(\rm B_1(\frac{3}{2}))$    & $\bstar\pi$, $\rm B\pi$,                     \\
         &                &                            & & &          $\bstarpipi$, $\bpipi$   \\
\hline
\end{tabular}
\caption{\label{t:fitconstraints} Constraints on production rates, masses and widths used in the
fit to the total $\rm B\pi$ mass spectrum.
}
\end{center} 
\end{table}

\item Only the decay modes listed in Table~\ref{t:fitconstraints} are
taken into account. We explicitly allow the decay via two pions to
$\bstar$ and $\rm B$ for all $\bstarj$ states. For each $\bstarj$
state, we set $\rm BR(\bstarj\rightarrow\bstarpipi) = BR(\bstarj
\rightarrow \bpipi)$ and we assume the same branching ratio $\brpipi$
for all $\bstarj$ states. For the $\Btwo$, we set $\rm
BR(\Btwo\rightarrow\bstar\pi) = BR(\Btwo\rightarrow B\pi)$.

\item The fraction of narrow states $\rm f_{narrow} = (f(b\rightarrow
\Bna)+f(b\rightarrow \Btwo)) / f(b\rightarrow\bstarj)$ is fixed to
0.6. This number is the average of 1/2, 2/3 and 2/3 corresponding to
production rates of narrow $\bstarj$ according to state counting,
total spin counting and light quark spin counting, respectively. To
justify this constraint, a fit to the total $\rm B\pi$ mass spectrum
is performed with $\rm f_{narrow}$ as an additional free parameter.

\item The mass splitting between $\rm B$ and $\bstar$ is fixed to the
current world average~\cite{pdg}.

\item Each of the five single pion decay modes is represented by a
Breit-Wigner function convoluted with the $M_{\rm B\pi}$ dependent
resolution function explained in Section~\ref{ss:bpimspectrum} and
shown in Figure~\ref{f:bpireso}. To take into account the phase-space
suppression at threshold, asymmetric Breit-Wigner functions with the
threshold factor~\cite{pythia}
\begin{equation}
f_{\rm threshold}(M) =
\sqrt { \left(1-\frac{M_{\rm B}^2}{M^2}-\frac{M_\pi^2}{M^2}\right)^2 -
4\cdot\frac{M_{\rm B}^2 M_\pi^2}{M^4}}
\end{equation}
are used instead of the symmetric Breit-Wigner functions for the broad
states.

\item For the di-pion transitions, the signal shape including the
detector resolution is taken from simulated $\bstarj \rightarrow \rm
B^{(*)}\pi\pi$ decays. Simple Gaussians truncated at threshold give a
good description of the simulated satellite peaks. The mean of the
Gaussian depends linearly on the mass difference between $\bstarj$ and
B ground state. The width of the Gaussian is also a function of this
mass splitting and depends on the width of the $\bstarj$ state. The
functions to parameterise the mean and width of the Gaussians are
taken from the simulation. No attempt is made to implement different
signal shapes for decays where the two pions form an intermediate
resonance or for cascade transitions from high mass $\bstarj$ states
via low mass $\bstarj$ states to the ground states $\bstar$ and B
\footnote{Strong decays within the $\bstarj$ multiplet are allowed if
the mass splitting within the multiplet or some of the widths are
larger than the pion mass.}.

\end{itemize}

The different reconstruction efficiencies for $\bstarjtobs$ and
$\bstarjtob$ decays of the $\bstar$-enriched and $\bstar$-depleted
signals are taken from Monte Carlo (see Table~\ref{t:effy}). As the
specific peaks that make up the total $\bstarj$ signal have different
sizes in the $\bstar$-enriched and $\bstar$-depleted mass
distributions, different enhancements or depletions are expected for
specific regions of the $\bpipm$ mass spectra according to the
assumptions on the nature of the $\bstarj$ signal. Thus the
simultaneous fit provides a non-trivial consistency check of the
$\brsb$ result of Section~\ref{ss:modindbrs} and of some of the
constraints used in the fit.

The implementation of $\brs$ as a fit parameter causes some
complications. Note that the $\brsb$ result of
Section~\ref{ss:modindbrs} includes transitions of narrow $\bstarj$
and broad $\bstarj$ via emission of one pion (and perhaps other decay
products). The branching fraction to the $\bstar$ ground state might
be different for narrow and broad states and also different for
$\bstar\pi$ and $\bstar\pi\pi$ final states. Whereas the result
obtained from Equation~\ref{formula:brs} is the average of the natural
composition of the different decay modes, four different parameters
have to be considered for this fit: ${\rm BR}(\bstarj_{\rm narrow}
\rightarrow {\rm B}^{*} \pi)$, ${\rm BR}(\bstarj_{\rm broad}
\rightarrow {\rm B}^{*} \pi)$, ${\rm BR}(\bstarj_{\rm narrow}
\rightarrow {\rm B}^{*} \pi\pi)$ and ${\rm BR}(\bstarj_{\rm broad}
\rightarrow {\rm B}^{*} \pi\pi)$. Monte Carlo studies indicate that
the sensitivity of the fit to ${\rm BR}(\bstarj_{\rm narrow}
\rightarrow {\rm B}^{*} \pi\pi)$ and ${\rm BR}(\bstarj_{\rm broad}
\rightarrow {\rm B}^{*} \pi\pi)$ is negligible since shape and
position of the corresponding peaks in the $\bpipm$ mass spectra are
almost the same for the $\bstar\pi\pi$ and $\rm B\pi\pi$ final
states. The sensitivity to ${\rm BR}(\bstarj_{\rm broad} \rightarrow
{\rm B}^{*} \pi)$ is also smaller than the corresponding sensitivity
to ${\rm BR}(\bstarj_{\rm narrow} \rightarrow {\rm B}^{*} \pi)$ since
a large width but a comparable intra-doublet mass splitting of the
$\bstarj$ states makes a separation of the decay modes of broad
$\bstarj$ to $\bstar\pi$ and $\rm B\pi$ difficult. To reduce the
number of fit parameters and to keep the correlations between the fit
parameters small, ${\rm BR}(\bstarj_{\rm narrow} \rightarrow {\rm
B}^{*} \pi\pi)/ ({\rm BR}(\bstarj_{\rm narrow} \rightarrow {\rm B}^{*}
\pi\pi) + {\rm BR}(\bstarj_{\rm narrow} \rightarrow {\rm B}\pi\pi))$
and ${\rm BR}(\bstarj_{\rm broad} \rightarrow {\rm B}^{*} \pi\pi)/
({\rm BR}(\bstarj_{\rm broad} \rightarrow {\rm B}^{*} \pi\pi) + {\rm
BR}(\bstarj_{\rm broad} \rightarrow {\rm B}\pi\pi))$ are fixed to 0.5
in the fit.  Furthermore, we require ${\rm BR}(\bstarj_{\rm narrow}
\rightarrow {\rm B}^{*} \pi) = \frac{3}{2}\cdot{\rm BR}(\bstarj_{\rm
broad} \rightarrow {\rm B}^{*} \pi)$ and the fit parameter $\brs$ is
the weighted mean of both numbers according to the production rates of
broad and narrow $\bstarj$. The factor $\frac{3}{2}$ is based on the
assumption of the same production rates for states within the same
doublet and $\rm BR(\Btwo\rightarrow\bstar\pi) /
(BR(\Btwo\rightarrow\bstar\pi) + BR(\Btwo\rightarrow B\pi)) = 0.5$.
The fit constraints explained in this paragraph are listed in
Table~\ref{t:fitconstraints2}.
\begin{table}[bth]
\begin{center}
\begin{tabular}{|c|c|}
\hline
 fit constraint                                                          &
 value                                                                   \\
\hline
$\frac{{\rm BR}({\rm B}^*_{J_{\rm narrow}} \rightarrow {\rm B}^{*} \pi\pi)}
{({\rm BR}({\rm B}^*_{J_{\rm narrow}} \rightarrow {\rm B}^{*} \pi\pi) +
 {\rm BR}({\rm B}^*_{J_{\rm narrow}} \rightarrow {\rm B}\pi\pi))}$       &
$\frac{1}{2}\rule[-3.0ex]{0cm}{7.0ex}$                                   \\
$\frac{{\rm BR}({\rm B}^*_{J_{\rm broad}} \rightarrow {\rm B}^{*} \pi\pi)}
{({\rm BR}({\rm B}^*_{J_{\rm broad}} \rightarrow {\rm B}^{*} \pi\pi) +
 {\rm BR}({\rm B}^*_{J_{\rm broad}} \rightarrow {\rm B}\pi\pi))}$        &
$\frac{1}{2}\rule[-3.0ex]{0cm}{7.0ex}$                                   \\
$\frac{{\rm BR}({\rm B}^*_{J_{\rm narrow}} \rightarrow {\rm B}^{*} \pi)}
{{\rm BR}({\rm B}^*_{J_{\rm broad}}  \rightarrow {\rm B}^{*} \pi)}$      &
$\frac{3}{2}\rule[-3.0ex]{0cm}{7.0ex}$                                   \\
\hline
\end{tabular}
\caption{\label{t:fitconstraints2} Additional fit constraints used in
the simultaneous fit.}
\end{center} 
\end{table}

With the constraints discussed above (see also
Tables~\ref{t:fitconstraints} and~\ref{t:fitconstraints2}) the
remaining free parameters are: sum of the number of entries of the
$\bstar$-enriched and the $\bstar$-depleted signal, $M(\Bna)$,
$\Gamma(\Bna)$, $M(\Bnull)$, $\Gamma(\Bnull)$, $\brpipi$ and $\brs$.
A fit~\cite{minuit} is performed to the simulated $\bstarj$ signals
using the full Monte Carlo statistics. All fit results lie within
$1\sigma$ of the Monte Carlo input value. We fit the $\bpipm$ mass
spectra of Figures~\ref{f:bsenr}c and \ref{f:bsdep}c
simultaneously. The least squares fit is performed in the $\bpi$ mass
region of $5.40-6.10\;\GeVcc$ with a bin width of $20\;\MeVcc$. The
fit results are
\begin{eqnarray*}
 M(\Bna)        & = & (\:\mb^{\:+\:\dmbplus}_{\:-\:\dmbminus}\:)    \;\GeVcc   \\
 \Gamma(\Bna)   & = & (\:\gb^{\:+\:\dgbstatplus}_{\:-\:\dgbstatminus}\:)             \;\MeVcc   \\
 M(\Bnull)      & = & (\:5.839^{\:+\:0.013}_{\:-\:0.014}\:)    \;\GeVcc \hspace*{22.0mm} (\star)  \\
 \Gamma(\Bnull) & = & (\:129^{\:+\:27}_{\:-\:23}\:)            \;\MeVcc \hspace*{28.45mm} (\star)  \\
 \brpipi        & = & 0.245^{\:+\:0.027}_{\:-\:0.028}                   \hspace*{41.3mm} (\star)  \\
 \brs           & = & \bra^{\:+\:\dbrastatplus}_{\:-\:\dbrastatminus}                        \\
\end{eqnarray*}
where the errors are of statistical origin only. The fit probability
is 65\% and the result is presented in Figures~\ref{f:bsenrfit} and
\ref{f:bsdepfit}. The fit results are in agreement with the
interpretation of the $\bstarj$ signals of Figure~\ref{f:pure} given
in Section~\ref{ss:modindbrs}. Numbers labelled with $(\star)$ should
be taken with care because of large systematic errors. The robustness
of the fit results will be discussed in Section~\ref{s:sys}.

Note that the statistical error of the $\brsb$ measurement presented
in Section~\ref{ss:modindbrs} includes the errors arising from the
sideband normalisation. The statistical errors of the fit results on
the other hand do not include the sideband normalisation errors. The
latter will be discussed in Section~\ref{s:sys}. The $\brs$ result of
the fit does not include decays to $\bstar$ via di-pion
emission. The $\brpipi$ result is corrected by a factor of 0.75 to
account for double counting of the $\rm B^{(*)}\pi^+\pi^-$ final
state. In Table~\ref{t:correlation} the correlations between all fit
parameters are shown. Systematic uncertainties of the fit results are
discussed in detail in Section~\ref{s:sys}.

\begin{table}[!htb]
  \begin{center} \begin{tabular}{|l|rrrrrrr|} \hline parameter &
    $N(\bstarj)$ & $M(\Bone)$ & $\Gamma(\Bone)$ & $M(\Bnull)$ &
    $\Gamma(\Bnull)$ & \scriptsize $\brpipi$ & \scriptsize $\brs$ \\
    \hline \hline $N(\bstarj)$ & 1.000 & 0.078 & 0.349 &-0.108 & 0.418
    &-0.088 & 0.114\\ $M(\Bna)$ & 0.078 & 1.000 & 0.394 & 0.028
    &-0.032 & 0.067 & 0.731\\ $\Gamma(\Bna)$ & 0.349 & 0.394 & 1.000
    &-0.675 & 0.355 &-0.764 & 0.791\\ $M(\Bnull)$ &-0.108 & 0.028
    &-0.675 & 1.000 &-0.313 & 0.741 &-0.380\\ $\Gamma(\Bnull)$ & 0.418
    &-0.032 & 0.355 &-0.313 & 1.000 &-0.473 &-0.036\\ \footnotesize
    $\brpipi$&-0.088 & 0.067 &-0.764 & 0.741 &-0.473 & 1.000 &-0.437\\
    \footnotesize $\brs$ & 0.114 & 0.731 & 0.791 &-0.380 &-0.036
    &-0.437 & 1.000\\ \hline \end{tabular} \end{center}
    \caption{Table
    of correlations of all free parameters in the final fit to the
    data mass spectra of the $\bstar$-enriched and $\bstar$-depleted
    samples.  \label{t:correlation}}
\end{table}
%
%
%
%
%
%
%
%
%
%
\section{Systematic uncertainties}\label{s:sys}

In the following sections, the determination of the systematic
uncertainties is presented separately for the model-independent
$\brsb$ measurement and for the results obtained from the simultaneous
fit to the $\bpipm$ mass spectra with different $\bstar$ content.

\boldmath
\subsection{Systematic error on the $\brsb$ measurement}\label{ss:modindsys}
\unboldmath

For the $\brsb$ measurement, the dominant sources of systematic error
are uncertainties in the efficiency ratios, the modelling of the
combinatorial $\bpipm$ background and systematic errors on the
sideband normalisation of the $\bstar$-enriched and $\bstar$-depleted
samples. Each contribution to the total error on $\brsb$ is listed in
Table~\ref{t:syserr}.

\boldmath
\subsubsection{Reconstruction efficiencies}\label{ss:sysphot}
\unboldmath

Monte Carlo simulations are used to calculate the efficiency ratios
$e$, $e^*$ and $e_0$. The systematic errors on these ratios are
dominated by uncertainties in the photon reconstruction. The
simulation is checked against data using known properties of $\bstar$
and $\pi^0$. The latter are formed by a pairwise combination of two
good conversion candidates or one good conversion and one good
calorimeter candidate assigned to the same B candidate.

\begin{itemize}

  \item We perform direct checks of the photon reconstruction
  efficiencies: The yields of the Monte Carlo $\bstar$ and $\pi^0$
  mass peaks are consistent with the results observed in data with the
  simulated $\bstar$ and $\pi^0$ production rates being in agreement
  with earlier measurements~\cite{pdg}. To account for the statistical
  error of the number of $\bstar$ and $\pi^0$ peak entries and for
  possible uncertainties in the simulated production rates, the
  calculation of efficiency ratios is repeated on Monte Carlo with the
  reconstruction efficiency of conversion (calorimeter) photons in the
  decay $\bstar\rightarrow {\rm B}\gamma$ changed by $\pm 10\%$ ($^{+
  15}_{-10}\%$). The variation of $+15\%$ reflects a small discrepancy
  observed in the simulated and measured $\bstar$ yields of the
  calorimeter sample.

  \item In a $\bstarj$ decay, the helicity angle $\theta^*$ is the
  angle between the signal pion momentum measured in the $\bstarj$
  rest frame and the momentum of the $\bstarj$ in the lab frame.  As
  the signal pion selection acceptance depends on $\cos\theta^*$, the
  $\bstarj$ efficiency is sensitive to the shape of $\cos\theta^*$.
  The distributions of helicity angle for $\Bnull$ and $\Bbr$ decays
  are assumed to be flat (S-wave transitions) and according to
  \cite{falkpeskin} the $\Bna$ and the $\Btwo$ are expected to have
  the same $\cos\theta^*$ distribution for any initial b polarisation:
  \begin{equation}
  \frac{1}{\Gamma} \frac{{\rm d}\Gamma}{{\rm d}\cos\theta^*} (\Bna
  ,\Btwo \rightarrow {\rm B},\bstar \pi ) = \frac{1}{4}
  \left(1+3\cos^2 \theta^*-6w_{3/2}
  (\cos^2\theta^*-\frac{1}{3})\right)
  \end{equation}
  where $w_{3/2}$ is the probability that fragmentation leads to a
  state with the maximum helicity value of 3/2 for the light degrees
  of freedom. The Monte Carlo $\cos\theta^*$ distributions of $\Bna$
  and $\Btwo$ have been reweighted to cover the whole range
  $w_{3/2}=0-1$.

  \item The number of good calorimeter photon candidates $N_{\gamma
  _{\rm ECAL}}$ (Figure~\ref{f:ncand}b) is not modelled well in the
  simulation.  Therefore, the Monte Carlo distribution is reweighted
  to the corresponding data distribution. The reweighting clearly
  improves the general agreement between data and Monte Carlo and has
  an impact on the efficiency rations $e$, $e^*$ and $e_0$. The
  central value of $\brsb$ changes by -0.059 due to the
  reweighting. To quantify the uncertainty in the reweighting
  procedure we take half of the total change of the central value as
  the error on $\brsb$.

  \item The mass dependent efficiency correction to the $\bstarj$
  signal for both the $\bstar$-enriched and $\bstar$-depleted samples
  produces a deviation in $\brsb$ relative to the result without mass
  dependent efficiency correction. To account for any mismodelling in
  the simulated $\bstarj$ masses and the simulated mass dependence of
  the efficiency, half of this deviation is taken as the systematic
  error.

  \item The calculated efficiency ratios are uncertain due to limited
  Monte Carlo statistics.

  \item We check the calibration of the photon energy measurement by
  comparing the measured and simulated shapes and peak positions of
  both the $\bstar$ and the $\pi^0$. There is agreement within the
  statistical errors. This results in small (negligible) errors on the
  efficiencies of the calorimeter (conversion) sample.

\end{itemize}

\boldmath
\subsubsection{Background related uncertainties}\label{ss:syspion}
\unboldmath

Uncertainties in the shape of the simulated background have an impact
on the number of signal candidates $N_{\rm E}$ and $N_{\rm D}$. Since
the combinatorial backgrounds in both the $\bstar$-enriched and the
$\bstar$-depleted samples are affected by systematic shifts in a
similar way, the measurement is rather robust against possible
uncertainties in the Monte Carlo background simulation. For the
determination of systematic errors, the simulated $\rm B\pi$
background is varied using the methods described below. For each
variation, the Monte Carlo background is normalised and subtracted
from the data and the number of signal entries $N_E$ and $N_D$ are
counted.

\begin{itemize}

 \item Data test samples are developed in which individual background
 sources are substantially enhanced to study the $\bpipm$
 combinatorial background. The background is divided into three
 different classes: tracks combined with mistagged B candidates in
 light and charm quark events (udsc flavour), b hadron decay tracks
 combined with true b hadrons (b hadron decay) and b fragmentation
 tracks combined with true b hadrons (b fragmentation). The selection
 criteria for each test sample are chosen to cover a large fraction of
 the kinematic region of the signal pion selection. A purity of at
 least 90\% for the background source under study and a $\bstarj$
 signal fraction smaller than 1.5\% is obtained by inverting cuts in
 the original $\bstarj$ selection. For the light and charm quark
 background, also a sample of ${\rm D}^{*+}$ candidates reconstructed
 as described in~\cite{ctol} is used as a qualitative cross check.
 For each test sample, the $\bpipm$ invariant mass distribution
 observed in data is compared with the Monte Carlo distribution
 normalised to the same number of selected B candidates. The mass
 distributions and their bin-by-bin ratios data/Monte Carlo are shown
 in Figure~\ref{f:bgcorr}. The different ratios are fitted with simple
 polynomials. The latter are used to correct the shape of the original
 Monte Carlo $\bpipm$ mass distributions for each background source
 separately. The systematic uncertainty on each background source is
 given by the difference in $\brsb$ between using the corrected and
 the uncorrected shape of the $\bpipm$ mass distribution.

 \item The composition of the $\bpipm$ background, as seen in the
 Monte Carlo after the corrections have been applied, is varied for
 each source. The fraction of each of the three background sources
 described earlier is varied by $\pm20\%$.

 \item The Peterson fragmentation parameter $\epsilon_{\rm b}$ has
 been varied in the range $0.0028-0.0057$ to cover uncertainties in
 the average fraction of the beam energy carried by the weakly
 decaying b hadron, $\meanxe$, and in the shape of the fragmentation
 function. This variation causes a minor change in the $\bpipm$
 background shape. The effect on $\brsb$ is smaller than 0.002.
  
 \item The average charged multiplicity of weakly decaying b hadrons
 (including $\rm K^0$ and $\Lambda$ decay products) is varied in the
 range $5.375-5.865$ in the simulation (see \cite{nb}). The observed
 effect on the $\rm B\pi^\pm$ background shape results in a negligible
 change of $\brsb$.

\end{itemize}

\subsubsection{$\!$Other sources of systematic uncertainties and
            consistency checks}\label{ss:othersys}

The following systematic studies have been performed in addition to
the studies described in Sections~\ref{ss:sysphot} and \ref{ss:syspion}.
\renewcommand{\baselinestretch}{1.0}
\begin{table}[tb]
\centering
\begin{tabular}{|l|c|c|}  
\hline
source                                        &  range           &  $\Delta (\brsb)$  \\ \hline\hline
$\bstar$ efficiency (ECAL) variation          &  $[0.90,1.15]$   &  $+0.036$ $-0.047$ \\
$\bstar$ efficiency (\gee ) variation         &  $[0.90,1.10]$   &  $+0.019$ $-0.018$ \\
$\cos \theta^*$ dependency                    & $w_{3/2}\in[0,1]$&  $+0.040$ $-0.034$ \\
reweighting of $N_{\gamma_{\rm ECAL}}$        &                  &  $\pm 0.030$       \\
$M_{\rm B\pi}$ dependence of $\bstarj$ efficiency &              &  $\pm 0.018$       \\
statistical error on efficiency ratios        &  $\approx 1\%$   &  $\pm 0.018$       \\
reconstructed $\bstar$ mass (ECAL)            &  $\pm 2\;\MeVcc$ &  $+0.007$ $-0.005$ \\
\hline
relative composition of background sources    & $\pm 20\%$       &  $+0.027$ $-0.037$ \\
B decay tracks background modelling           &  corr. on/off    &  $\pm 0.017$       \\
b fragmentation tracks background modelling   &  corr. on/off    &  $\pm 0.005$       \\
udsc tracks background modelling              &  corr. on/off    &  $\pm 0.002$       \\
\hline
sideband range variation                      & $\pm100\;\MeVcc$ &  $+0.076$ $-0.057$ \\
variation of cuts on $\bsprob$                &                  &  $+0.030$ $-0.043$ \\
${\rm B}^{*}_{sJ}$ reflections                &                  &  $+0.006$ $-0.026$ \\
${\rm B}^{(*)\prime}$ reflections             &                  &  $+0.000$ $-0.017$ \\
track parameter resolution variation          & $\pm 10\%$       &  $<0.010$          \\
\hline\hline                                                   
total                                         &                  &  $+0.12$ $-0.12$ \\
\hline
\end{tabular}
\renewcommand{\baselinestretch}{1.0}
\caption{\label{t:syserr} Systematic errors of the $\brsb$
measurement.  Detailed information for each uncertainty is given in
the text, as well as a discussion of uncertainties which are
negligible and thus excluded from this table.}
\end{table}

\begin{itemize}

 \item The range of the sideband used for the background normalisation
 is varied by $\pm 100\;\MeVcc$ on each side for both the
 $\bstar$-enriched and $\bstar$-depleted sample. This variation is
 motivated by the range and shape of the $\bstarj$ signal observed in
 Figure~\ref{f:bdsall}. The quadratic sum of the differences observed
 in the number of signal entries gives the largest error contribution
 to the systematic error of the $\brsb$ measurement.

 \item The cut on $\bsprob$ has been varied. All cut values producing
 a ratio of signal to noise ratios of the $\bstar$-enriched and
 $\bstar$-depleted samples between 0.9 and 1.1 are considered. The
 observed deviations in $\brsb$ do not exceed $^{+0.030}_{-0.043}$
 which is taken as the systematic error.

 \item The amount of ${\rm B}^{*}_{sJ}$ seen in the $\bstarj$ peaks is
 less than 4\% for a ${\rm B}^{*}_{sJ}$ production rate consistent
 with \cite{bdstar}. The branching ratio ${\rm BR} ({\rm B}^{*}_{sJ}
 \rightarrow {\rm B}^{*}{\rm K})$ is varied from 0.2 to 1.0 in the
 simulation.

 \item Contributions from radial excitations of B mesons decaying to
 ${\rm B}^{(*)}\pi$ or ${\rm B}^{(*)}\pi\pi$ may be present in the
 $\bstarj$ signal~\cite{bdstarl3,bstarprime}. Monte Carlo studies with
 simulated ${\rm B}^{\prime}\rightarrow {\rm B}\pi^+\pi^-$ and ${\rm
 B}^{*\prime} \rightarrow {\rm B}^*\pi^+\pi^-$ decays\footnote{using
 $M({\rm B^{\prime}})=5.883\;\GeVcc$ and $M({\rm B^{*\prime}})=
 5.898\;\GeVcc$ according to \cite{ebertfaustovgalkin} and in
 agreement with \cite{bstarprime}} indicate a contamination of the
 $\bstarj$ signal around $M_{\rm B\pi}=5.6\;\GeVcc$. With the total
 ${\rm B^{(*) \prime}}$ production rate observed in \cite{bstarprime},
 the $\bstarj$ peak of Figure~\ref{f:bdsall} does not contain more
 than 3\% ${\rm B}^{(*)\prime}$ transitions. The fraction of ${\rm
 B}^{\prime}$ in the simulated ${\rm B}^{(*)\prime}$ sample is varied
 from 0.3 to 0.7.

 \item To account for any uncertainties arising from a wrongly
 simulated tracking resolution, the reconstructed track parameters are
 smeared by $\pm 10\%$ in the Monte Carlo~\cite{rb}.

 \item The cuts of the signal pion selection have been varied.  No
 systematic deviations are observed.

 \item The whole analysis is repeated using conversion photons only
 and calorimeter photons only. The obtained $\brsb$ results and the
 $\bstarjtobsb$ and $\bstarjtobb$ mass distributions are in agreement
 with each other and with the total sample.

 \item Varying the cut on ${\cal B}_{\rm event}$ so that the b purity
 changes from 92\% to 98\% produces no systematic deviation in
 $\brsb$.

 \item A neural network has been trained to replace the weight
 $\bsprob$ obtained from the Monte Carlo purity parameterisation. The
 neural network output is strongly correlated with the weight
 $\bsprob$ and does not improve the $\bstar$/$\rm B$ separation.

\end{itemize}

All systematic errors considered for $\brsb$ are added up in
quadrature, resulting in a total error on $\brsb$ of $\pm 0.12$ (see
Table~\ref{t:syserr}).

\boldmath
\subsection{Systematic uncertainties on the results
            of the simultaneous fit to the $\bstar$-enriched and
            $\bstar$-depleted mass spectra}
\unboldmath

All sources of systematic error are varied in turn within the
estimated uncertainty range and the fit presented in
Section~\ref{ss:simulfit} is repeated. If not stated otherwise, the
observed deviation \wrt\ the original fit is taken as a systematic
error for each fit parameter. The total systematic error on each fit
parameter is the quadratic sum of all individual error
contributions. This procedure takes into account the correlations
between the fit parameters. All systematic uncertainties are listed in
Table~\ref{t:simultfitsys}.

\subsubsection{Variation of fit constraints}\label{sss:constraintsys}

The fit constraints are varied according to the present knowledge of
$\rm D^*_J$ properties \cite{pdg} and theoretical considerations
\cite{ehq,ebertfaustovgalkin,godfreykokoski,manyothers,isgur,aksel}:

\begin{itemize}

 \item ${\rm BR} ( \bstarj_{\rm broad} \rightarrow {\rm B}^{*}\pi)$
 was fixed to $\frac{2}{3}\cdot{\rm BR}(\bstarj_{\rm narrow}
 \rightarrow {\rm B}^{*}\pi)$. This constraint is changed to ${\rm
 BR}(\bstarj_{\rm broad} \rightarrow {\rm B}^{*}\pi)= {\rm
 BR}(\bstarj_{\rm narrow} \rightarrow {\rm B}^{*}\pi)$ and the
 corresponding deviations of the fit parameters are taken as
 systematic errors.

 \item The ratio of the widths of partners of the same doublet was
 fixed to $\Gamma(\Btwo) / \Gamma(\Bna) = 1.25$ and $\Gamma(\Bbr) /
 \Gamma(\Bnull) = 1.00$. We allow a variation of the ratio of widths
 of 1.0-1.4 for the narrow states and 0.7-1.4 for the broad states.

 \item The relative production rates $\rm
 f(b\rightarrow\Bnull):f(b\rightarrow\Bbr):f(b\rightarrow\Bna):f(b\rightarrow\Btwo)$
 were fixed to $2:2:3:3$. We vary the production rate ratio $\rm
 f(b\rightarrow\Bna) / (f(b\rightarrow\Bna)+f(b\rightarrow\Btwo))$
 in the range 0.375-0.600, $\rm f(b\rightarrow\Bnull) /
 (f(b\rightarrow\Bnull)+f(b\rightarrow\Bbr)$ in the range 0.250-0.600
 and $\rm f_{narrow}$ in the range 0.5-0.75. These variations cover
 the different production rate estimates of state counting (1:1:1:1)
 to total spin counting (1:3:3:5).

 \item The constraint $\rm BR(\Btwo\rightarrow\bstar\pi) =
 BR(\Btwo\rightarrow B\pi)$ was used. We allow a variation of $\rm
 BR(\Btwo\rightarrow\bstar\pi) / (BR(\Btwo\rightarrow\bstar\pi) +
 BR(\Btwo\rightarrow B\pi)) = 0.3-0.7$.

 \item Whereas the mass splitting between the $j_q=1/2$ and $j_q=3/2$
 doublets was free in the fit, the mass splittings within the doublets
 were fixed. We allow a variation of $M(\Btwo)-M(\Bna) =
 (5-20)\;\MeVcc$ and $M(\Bbr)-M(\Bnull) = (0-50)\;\MeVcc$.

 \item The ratio ${\rm BR} (\bstarj \rightarrow {\rm B}^{*} \pi\pi)/
 ({\rm BR} (\bstarj \rightarrow {\rm B}^{*} \pi\pi)+ {\rm BR} (\bstarj
 \rightarrow {\rm B} \pi\pi))$, which was fixed to 0.5, is varied in
 the range 0.3-0.7. No significant deviation in the fit parameters is
 observed, since the position and shape of the peaks corresponding to
 the two final states are very similar.

\end{itemize}

\subsubsection{Reconstruction efficiencies}\label{sss:effy}

The $\bstarj$ reconstruction efficiency is a function of the
reconstructed $\rm B\pi$ mass and the angular distribution of the
$\pi$. The efficiency is taken from the Monte Carlo and possible
problems with its simulation are taken into account:

\begin{itemize}

 \item The signal pion selection cuts have been varied. Whereas the
 Monte Carlo simulation describes the reconstruction efficiency well
 at high $\rm B\pi$ masses, this statement can not be proven for low
 $\rm B\pi$ masses. Therefore, we perform the fit to the $\bstarj$
 signal without an acceptance correction of the mass spectrum as shown
 in Figure~\ref{f:bdsall}b. Although the fit result has a low fit
 probability, we assign half of the total deviations of the fit
 parameters observed \wrt\ the original fit as systematic errors. This
 results in a large systematic error on $\brpipi$.

 \item Systematic uncertainties arising from the photon reconstruction
 as well as from the unknown $\cos\theta^*$ distributions are
 determined using the methods presented in~\ref{ss:sysphot}.

\end{itemize}

\subsubsection{Background related uncertainties}\label{sss:bg}

\renewcommand{\baselinestretch}{1.3}
\begin{table}
{\centering
{\small
\begin{tabular}{|l|c|c|c|c|c|c|}  
\hline
 & \raisebox{-0.5ex}[-0.5ex]{$\Delta {\rm BR}$} & & & & &
   \raisebox{-0.5ex}[-0.5ex]{$\frac{1}{0.75}\cdot\Delta {\rm BR}$} \\

 \raisebox{2.0ex}[-2.0ex]{source}  &
 \raisebox{0.5ex}[-0.5ex]{\footnotesize $\!(\bstarj \rightarrow {\rm B}^{*} \pi)\!\!$} &
 \raisebox{2.0ex}[-2.0ex]{\small $\!\Delta M(\rm B_1(\frac{3}{2}))\!$}   &
 \raisebox{2.0ex}[-2.0ex]{\small $\!\!\Delta\Gamma(\rm B_1(\frac{3}{2}))\!\!$}&
 \raisebox{2.0ex}[-2.0ex]{$\Delta M(\Bnull)$} &
 \raisebox{2.0ex}[-2.0ex]{$\Delta\Gamma(\Bnull)$} &
 \raisebox{0.5ex}[-0.5ex]{\scriptsize $\!\!(\bstarj \rightarrow {\rm B}^{(*)} \pi\pi)\!\!$}  \\
\hline\hline 
\footnotesize ${\rm BR}\:\!(\;\!\bstarj_
   {\rm broad} \rightarrow {\rm B}^{*}\pi)$  & $+0.059$              & $-0.0015$               & $-4.8$            & $+0.0097$              & $+1$            & $+0.011$                 \\
\footnotesize $\Gamma(\Btwo)/\Gamma(\Bna)$    & $^{+0.014}_{-0.009}$ & $^{+0.0008}_{-0.0013}$  & $^{+4.1}_{-1.9}$  & $^{+0.0001}_{-0.0029}$ & $^{+2}_{-0}$   & $^{+0.007}_{-0.016}$     \\
\footnotesize $\Gamma(\Bbr)/\Gamma(\Bnull)$   & $+0.001$             & $+0.0004$               & $+0.5$            & $-0.0030$              & $-20$           & $-0.004$                 \\
\footnotesize $\rm f_{narrow}$                & $^{+0.049}_{-0.028}$ & $^{+0.0001}_{-0.0002}$  & $^{+7.9}_{-8.9}$  & $^{+0.0144}_{-0.0199}$ & $\pm 12$        & $^{+0.001}_{-0.008}$     \\
\footnotesize prod. rate $\Bna$ vs. $\Btwo$   & $\pm 0.012$          & $^{+0.0015}_{-0.0021}$  & $\pm 1.2$         & $^{+0.0000}_{-0.0022}$ & $\pm 1$         & $^{+0.003}_{-0.006}$     \\
\footnotesize prod. rate $\Bnull$ vs. $\Bbr$  & $^{+0.004}_{-0.003}$ & $^{+0.0001}_{-0.0002}$  & $\pm 0.0$         & $^{+0.0060}_{-0.0054}$ & $^{+1}_{-2}$   & $^{+0.003}_{-0.002}$     \\
\footnotesize $\rm BR(\Btwo\rightarrow\bstar                           
                                \pi / B\pi)$  & $^{+0.013}_{-0.008}$ & $^{+0.0012}_{-0.0004}$  & $^{+2.4}_{-1.2}$  & $^{+0.0043}_{-0.0075}$ & $^{+3}_{-2}$   & $^{+0.012}_{-0.017}$     \\
\footnotesize $M(\Btwo)-M(\Bna)$              & $^{+0.003}_{-0.002}$ & $\pm 0.0024$            & $^{+2.4}_{-3.2}$  & $^{+0.0017}_{-0.0010}$ & $\pm 2$         & $^{+0.003}_{-0.002}$     \\
\footnotesize $M(\Bbr)-M(\Bnull)$             & $\pm 0.001$          & $\pm 0.0001$            & $^{+0.0}_{-0.2}$  & $^{+0.0119}_{-0.0200}$ & $^{+5}_{-6}$   & $^{+0.007}_{-0.012}$     \\
\hline
\footnotesize efficiency $f(M(\rm B\pi))$     & $\pm 0.039$             & $\pm 0.0018$               & $\pm 4.3$            & $\pm 0.0050$              & $\pm 1$            & $\pm 0.109$                 \\
\footnotesize reweighting of $N_{\gamma_{\rm ECAL}}$& $\pm 0.036$    & $\pm 0.0021$            & $\pm 7.4$         & $\pm 0.0025$           & $\pm 2$         & $\pm 0.021$              \\
\footnotesize $\bstar$ efficiency (ECAL)      & $^{+0.013}_{-0.023}$ & $^{+0.0007}_{-0.0005}$  & $^{+1.3}_{-2.3}$  & $^{+0.0004}_{-0.0016}$ & $+1$            & $\pm 0.003$              \\
\footnotesize efficiency $f(\cos \theta^{*})$ & $^{+0.002}_{-0.001}$ & $^{+0.0001}_{-0.0002}$  & $^{+4.1}_{-5.0}$  & $^{+0.0077}_{-0.0103}$ & $^{+6}_{-7}$   & $^{+0.002}_{-0.004}$     \\
\hline

\footnotesize sideband norm. $\bstar$-enr.    & $^{+0.125}_{-0.071}$ & $^{+0.0041}_{-0.0032}$  & $^{+15.0}_{-9.3}$ & $^{+0.0013}_{-0.0020}$ & $^{+3}_{-6}$   & $^{+0.017}_{-0.010}$     \\
\footnotesize sideband norm. $\bstar$-dep.    & $^{+0.116}_{-0.067}$ & $^{+0.0031}_{-0.0025}$  & $^{+9.6}_{-4.5}$  & $^{+0.0075}_{-0.0081}$ & $^{+4}_{-9}$   & $^{+0.014}_{-0.022}$     \\
\footnotesize Peterson fragmentation          & $^{+0.034}_{-0.049}$ & $^{+0.0010}_{-0.0008}$  & $^{+12.3}_{-6.4}$& $^{+0.0044}_{-0.0047}$ & $^{+35}_{-27}$ & $^{+0.046}_{-0.065}$     \\
\footnotesize background fit function         & $\pm 0.023$          & $\pm 0.0025$            & $\pm 0.6$         & $\pm 0.0055$           & $\pm 6$         & $\pm 0.005$              \\
\footnotesize sideband range variation        & $^{+0.032}_{-0.030}$ & $^{+0.0012}_{-0.0011}$  & $^{+3.9}_{-5.2}$  & $^{+0.0037}_{-0.0064}$ & $^{+5}_{-13}$  & $^{+0.005}_{-0.007}$     \\
\footnotesize B decay bg fraction             & $^{+0.037}_{-0.044}$ & $^{+0.0001}_{-0.0008}$  & $^{+4.7}_{-5.0}$  & $^{+0.0089}_{-0.0082}$ & $^{+4}_{-6}$   & $\pm 0.054$              \\
\footnotesize B decay bg modelling            & $\pm 0.016$           & $\pm 0.0012$           & $\pm 6.9$        & $\pm 0.0111$           & $\pm 25$        & $\pm 0.002$              \\
\footnotesize b fragm. bg fraction            & $^{+0.030}_{-0.028}$ & $^{+0.0003}_{-0.0004}$  & $^{+5.7}_{-4.4}$  & $^{+0.0060}_{-0.0084}$ & $^{+10}_{-8}$  & $^{+0.042}_{-0.061}$     \\
\footnotesize b fragm. bg modelling           & $\pm 0.025$           & $\pm 0.0002$           & $\pm 4.7$         & $\pm 0.0104$           & $\pm 33$        & $\pm 0.007$              \\
\footnotesize udsc bg fraction                & $\pm 0.023$          & $^{+0.0001}_{-0.0000}$  & $\pm 0.5$         & $\pm 0.0007$           & $\pm 0$         & $\pm 0.002$              \\
\hline
\footnotesize variation of bin width           & $^{+0.009}_{-0.005}$ & $^{+0.0010}_{-0.0008}$  & $^{+5.0}_{-3.5}$  & $^{+0.0044}_{-0.0076}$ & $^{+5}_{-16}$  & $^{+0.017}_{-0.033}$     \\
\footnotesize$\rm B^{(*)\prime}$ contamination& $^{+0.026}_{-0.028}$ & $^{+0.0003}_{-0.0002}$  & $^{+3.5}_{-3.7}$  & $^{+0.0055}_{-0.0044}$ & $^{+4}_{-5}$   & $\pm 0.003$              \\
\footnotesize $\bstarjs$ contamination        & $\pm 0.021$          & $^{+0.0002}_{-0.0004}$  & $\pm 2.7$         & $^{+0.0042}_{-0.0043}$ & $\pm 4$         & $^{+0.004}_{-0.005}$     \\
\footnotesize variation of fit range           & $\pm 0.002$          & $^{+0.0001}_{-0.0000}$  & $^{+2.0}_{-1.7}$  & $^{+0.0009}_{-0.0006}$ & $^{+11}_{-10}$ & $^{+0.010}_{-0.006}$     \\
\hline\hline
\normalsize \rule[-2.5mm]{0mm}{8mm} total    & \large $^{+\dbrasysplus}_{-\dbrasysminus}$    & $\pm \dmbsys$          &
\large $^{+\dgbsysplus}_{-\dgbsysminus}$  & \large  $^{+0.034}_{-0.042}$   & $\pm 63$        &
\large $^{+0.143}_{-0.161}$     \\
\hline\hline
\footnotesize \rule[-2mm]{0mm}{7mm} central value of fit result & $\bra$ & $\mb$ & $\gb$  & $5.839$ & $129$ & $0.327$ \\
\hline
\end{tabular}}}
\renewcommand{\baselinestretch}{1.0}
\caption{\label{t:simultfitsys} Systematic errors on the fit
parameters of the simultaneous fit. The numbers for $\Delta\brpipi$
have to be multiplied by 0.75 to account for double counting. The
total systematic error of each fit parameter is the quadratic sum of
the individual errors. Sources of systematic uncertainty with an error
of smaller than 10\% of the total systematic error for {\em each} fit
parameter are excluded from the table.}
\end{table}

\begin{itemize}

 \item The uncertainties due to the limited data statistics in the
 upper sideband regions of the $\bstar$-enriched and $\bstar$-depleted
 samples produce quite large error contributions on $\brs$. For the
 fit results, these error contributions are treated as systematic
 errors and have to be compared with the corresponding errors obtained
 for the model-independent $\brsb$ measurement.
While the sideband range used for the background normalisation is the
same for both measurements, the error of the $\brs$ fit result is
smaller because the fit makes explicit use of the shape and of the
composition of the $\bstarj$ mass distributions.

 \item The Peterson fragmentation parameter $\epsilon_{\rm b}$ and the
 charged particle multiplicity of weakly decaying b hadrons are
 varied in the simulation as described in Section~\ref{ss:syspion}.
 The multiplicity variation has no effect on the fit results.

 \item To evaluate the uncertainty due to the function used to fit the
 Monte Carlo background, we directly subtract the corrected Monte
 Carlo background histograms of the $\rm B\pi$ mass distributions from
 the corresponding data histograms.

 \item The range of the sideband used for the background normalisation
 is varied by $\pm 100\;\MeVcc$ on each side and the observed deviations
 are added in quadrature.

 \item For the determination of the shape and the composition of the
 $\bpi$ background, the same methods as presented in
 Section~\ref{ss:syspion} are used.

\end{itemize}

\boldmath
\subsubsection{Contamination of the $\bstarj$ signal, other uncertainties and cross checks}
\label{sss:contaminationsys}
\unboldmath

\begin{itemize}
 
 \item Contributions from higher orbital or radial excitations of B
 mesons decaying to ${\rm B}^{(*)}\pi$ or ${\rm B}^{(*)}\pi\pi$ may be
 contained in the signal peaks of the $\bstar$-enriched and the
 $\bstar$-depleted samples. The procedure described in
 Section~\ref{ss:sysphot} is repeated.

 \item Reflections from ${\rm B}^{*}_{sJ}$ decays influence the fit
 results since a small fraction of kaons are misidentified as
 pions. The $\bstarjs$ production rate is varied by $\pm 50\%$ to
 cover the experimental error~\cite{bdstar}.

 \item The range of the fit region is changed by $\pm 80\;\MeVcc$ on
 both sides.

 \item The whole analysis is performed using bin widths of 16 and
 $25\;\MeVcc$ instead of $20\;\MeVcc$.

 \item A fit to total $\bpipm$ mass spectrum produces results for
 $M(\Bna)$, $\Gamma(\Bna)$, $M(\Bnull)$, $\Gamma(\Bnull)$ and
 $\brpipi$ in good agreement with the fit results obtained from the
 mass spectra of the $\bstar$-enriched and $\bstar$-depleted
 samples. Furthermore, the fit to the total $\rm B\pi$ mass spectrum
 is repeated with $\rm f_{narrow}$ as an additional fit parameter. The
 result $\rm f_{narrow} = 0.76^{\:+\:0.11}_{\:-\:0.24 \;\rm stat}$ is
 in agreement with the fixed value of 0.6 used in the original fit.

 \item The number of $\bstarj$ signal entries observed in
 Figure~\ref{f:bdsall} and the Monte Carlo reconstruction efficiencies
 produce a total $\bstarj$ production rate consistent with our
 measurement presented in~\cite{bdstar}.

\end{itemize}

\boldmath
\subsubsection{Robustness of fit results}
\label{sss:robustness}
\unboldmath

Extensive systematic studies have been performed to test the stability
of the fit results. Since the systematic error on $\brpipi$ is large,
the existence of $\bstarj\to \rm B^{(*)}\pi\pi$ decays can be
questioned and the fit is repeated with $\brpipi=0$. Furthermore, the
functional form of the broad $\bstarj$ states, which is not precisely
known at $\rm B\pi$ threshold, has been varied assuming different
theoretical approaches. The fit probability for any fit with $\brpipi$
fixed to zero is always below 4\% (to be compared with a fit
probability of 65\% of the original fit). Depending on the functional
form of the broad $\bstarj$ states, the mass of the broad $\bstarj$
states lies below or above the $\Bna$ mass for $\brpipi=0$. Because of
this ambiguity, the widths and the mass of the $\Bnull$ as well as
$\brpipi$ are not quoted as robust fit results. The mass and the width
of the $\Bna$ as well as $\brs$ stay stable for all fits. The
described ambiguity is also observed in the corresponding fit to the
total $\rm B\pi$ mass spectrum. In comparison to other measurements,
especially~\cite{bdstarl3}, this is a rather conservative treatment of
the fit results. Experiments with comparable mass resolution are
expected to have similar problems in determining $\bstarj$ properties
from a fit to a single $\rm B\pi$ mass peak.

%
%
%
%
%
%
%
%
\section{Summary and conclusion}\label{s:sum}

We have analysed orbitally-excited mesons by forming combinations of
inclusively reconstructed B mesons and charged pions. A new way to
determine the combinatorial $\rm B\pi$ background using data test
samples while maintaining high statistics is presented. A high
statistics tag of the decay $\rm \bstar\rightarrow B\gamma$ is used to
obtain $\bstarj$ samples enriched or depleted in their $\bstar$
content. We present the first measurement of the branching ratio of
orbitally-excited B mesons decaying into $\bstar$. The result is
\[
\brsb = \brval^{\,\brstatplus}_{\,\brstatminus} \pm \brsystot \, ,
\]
where the first error is statistical and the second systematic. The
measurement does not depend on the shape of $\bpi$ mass distributions
or on any specific model. It is in agreement with theoretical
predictions and the measured $\bstar$ and $\bstarj$ production rates
at LEP.

Making further use of the $\bstar$ information, a simultaneous fit to
the $\bstarj$ mass spectra of samples enriched or depleted in their
$\bstar$ content is performed. In this fit, the masses, widths and
production rates of the $\Bbr$ and $\Btwo$ are constrained by the
corresponding properties of their doublet partners $\Bnull$ and
$\Bna$, respectively. The fit yields
\begin{eqnarray*}
 M(\Bna)      & = & (\:\mb ^{\:+\:\dmbplus                      }_{\:-\:\dmbminus}\pm \dmbsys \:)          \;\GeVcc \\
 \Gamma(\Bna) & = & (\:\gb ^{\:+\:\dgbstatplus\:+\:\dgbsysplus  }_{\:-\:\dgbstatminus\:-\:\dgbsysminus}\:) \;\MeVcc \\
 \brs         & = &  \bra  ^{\:+\:\dbrastatplus\:+\:\dbrasysplus}_{\:-\:\dbrastatminus\:-\:\dbrasysminus}\;.
\end{eqnarray*}
The first error indicates the statistical and the second error the
systematic uncertainty. The fit favours a contribution of
$\bstarj\to\rm B^{(*)}\pi\pi$ decays to the $\bstarj$ signal and a
mass of the broad $\bstarj$ states about $100\;\MeVcc$ above the
narrow $\bstarj$ states. Systematic uncertainties in the
reconstruction efficiency and the combinatorial background at low $\rm
B\pi$ masses together with the lack of knowledge of the exact
functional form of the broad $\bstarj$ states at $\rm B\pi$ threshold
do not allow an unambiguous determination of the width and mass of the
$\Bnull$ (or $\Bbr$). The fit results are in agreement with
predictions from several HQET models. The $M(\Bna)$ result agrees well
with a measurement of $M(\Btwo)$ \cite{bdsalephexcl}. On the other
hand, a recent $\bstarj$ analysis~\cite{bdstarl3} presented masses
that disagree with the results of this analysis and with
\cite{bdsalephexcl}. The measured value of $\brpipi= 0.25 \pm 0.02
^{\:+\:0.11}_{\:-\:0.12}$ is consistent with the range 0.1-0.2
predicted by theory~\cite{ehq} and in agreement with an experimental
result obtained from the reconstruction of the $\rm B^{(*)}\pi\pi$
final state~\cite{bstarprime}. The results for $\brsb$, $\brs$ and
$\brpipi$ are in good agreement with each other on the assumption that
the decay channels with $X\ne\pi$ contributing to $\brsb$ are small.

\bigskip\bigskip\bigskip
\appendix
\par
Acknowledgements:
\par
We particularly wish to thank the SL Division for the efficient operation
of the LEP accelerator at all energies
 and for their continuing close cooperation with
our experimental group.  We thank our colleagues from CEA, DAPNIA/SPP,
CE-Saclay for their efforts over the years on the time-of-flight and trigger
systems which we continue to use.  In addition to the support staff at our own
institutions we are pleased to acknowledge the  \\
Department of Energy, USA, \\
National Science Foundation, USA, \\
Particle Physics and Astronomy Research Council, UK, \\
Natural Sciences and Engineering Research Council, Canada, \\
Israel Science Foundation, administered by the Israel
Academy of Science and Humanities, \\
Minerva Gesellschaft, \\
Benoziyo Center for High Energy Physics,\\
Japanese Ministry of Education, Science and Culture (the
Monbusho) and a grant under the Monbusho International
Science Research Program,\\
Japanese Society for the Promotion of Science (JSPS),\\
German Israeli Bi-national Science Foundation (GIF), \\
Bundesministerium f\"ur Bildung und Forschung, Germany, \\
National Research Council of Canada, \\
Research Corporation, USA,\\
Hungarian Foundation for Scientific Research, OTKA T-029328, 
T023793 and OTKA F-023259.\\

%
%
%
%
%
%
%
%

%
%

\epostfigtwo{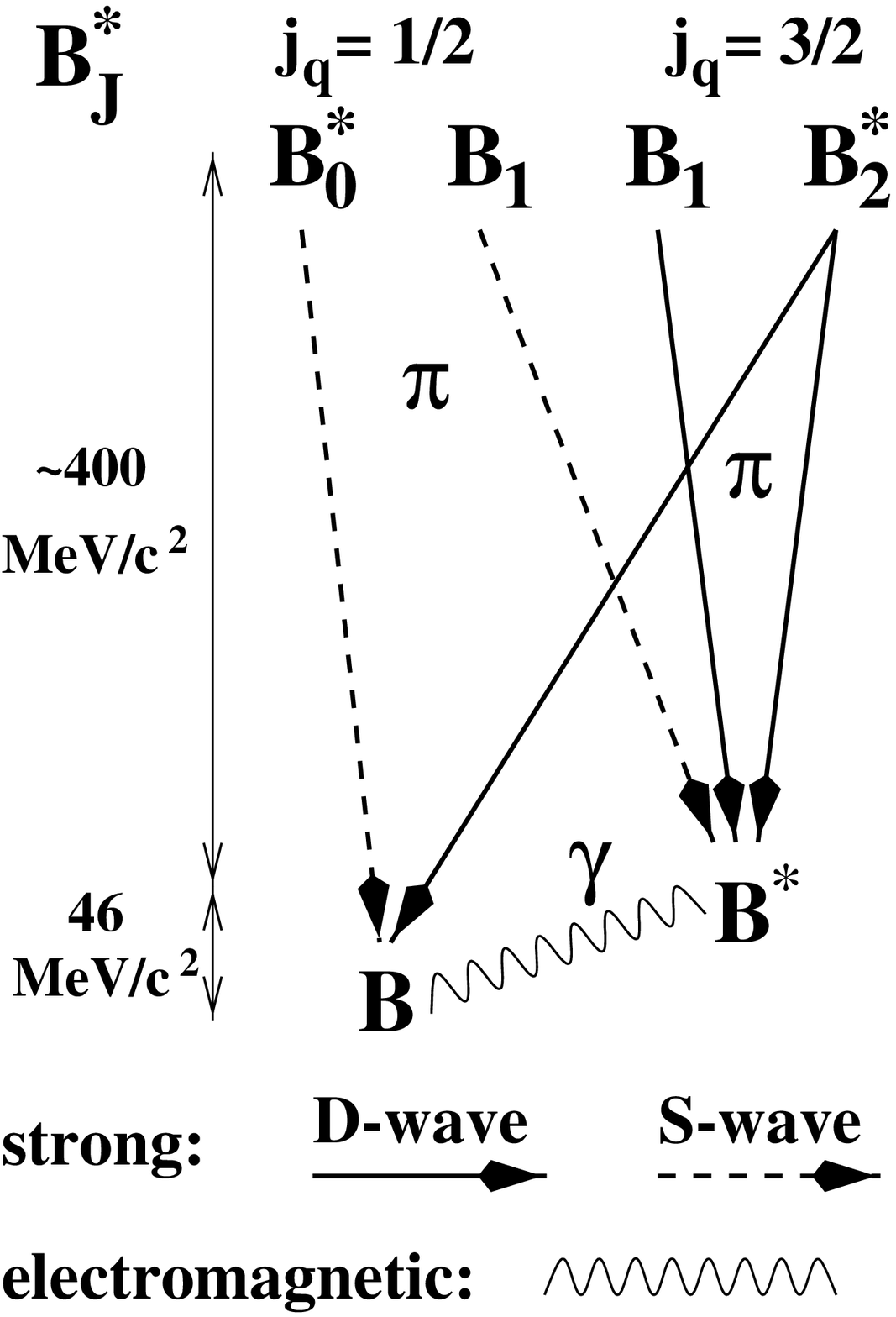}{f:transition}{ The four
  $\bstarj$ states and their dominant decays to the ground state
  doublet ($\rm B$, $\bstar$). Strong decays via single pion emission
  are indicated as solid (D-wave) and dashed (S-wave) lines.
  The $\bstar$ decays radiatively because of the small
  $\bstar$--$\rm B$ mass splitting.}

\epostfig{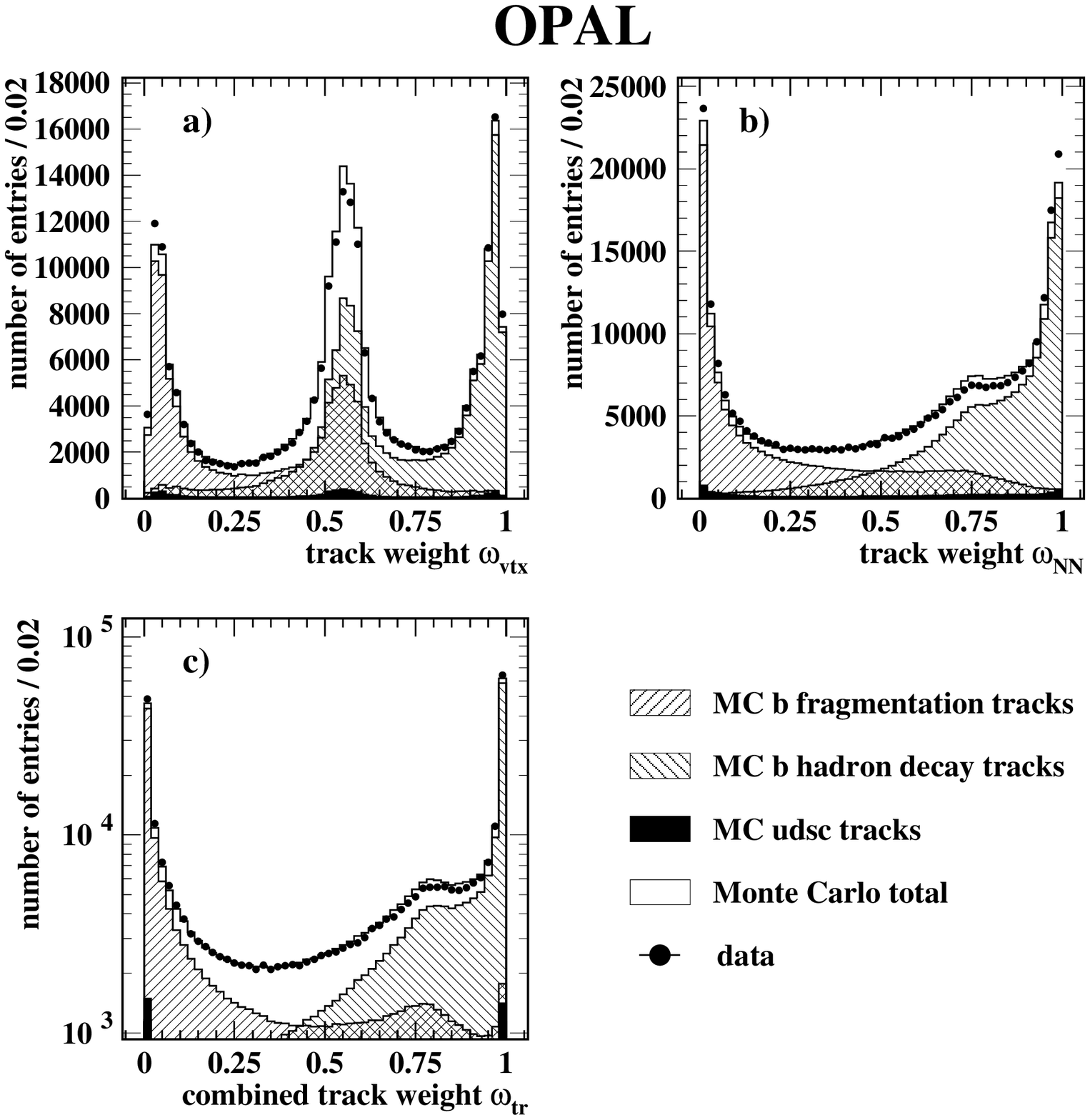}{f:trackweights}{a) The track weight
  $\omega_{\rm vtx}$ for all tracks in hemispheres with a good
  secondary vertex. The peaks near 0 and 1 correspond to tracks
  created by b fragmentation and b hadron decay tracks,
  respectively. The peak near 0.5 is produced by tracks which are not
  unambiguously assigned to the primary or the secondary vertex, as in
  the case of tracks matching both the primary and secondary vertex or
  matching no vertex at all. b) The track weight $\omega_{\rm NN}$ for
  tracks of all hemispheres (with or without a good secondary
  vertex). The separation power of $\omega_{\rm NN}$ is superior to
  the separation power of $\omega_{\rm vtx}$. c) The combined track
  weight $\omega_{\rm tr}$ calculated from $\omega_{\rm vtx}$ and
  $\omega_{\rm NN}$ for tracks of all hemispheres. Note that
  $\omega_{\rm tr}$ is shown on a logarithmic scale.}

\epostfig{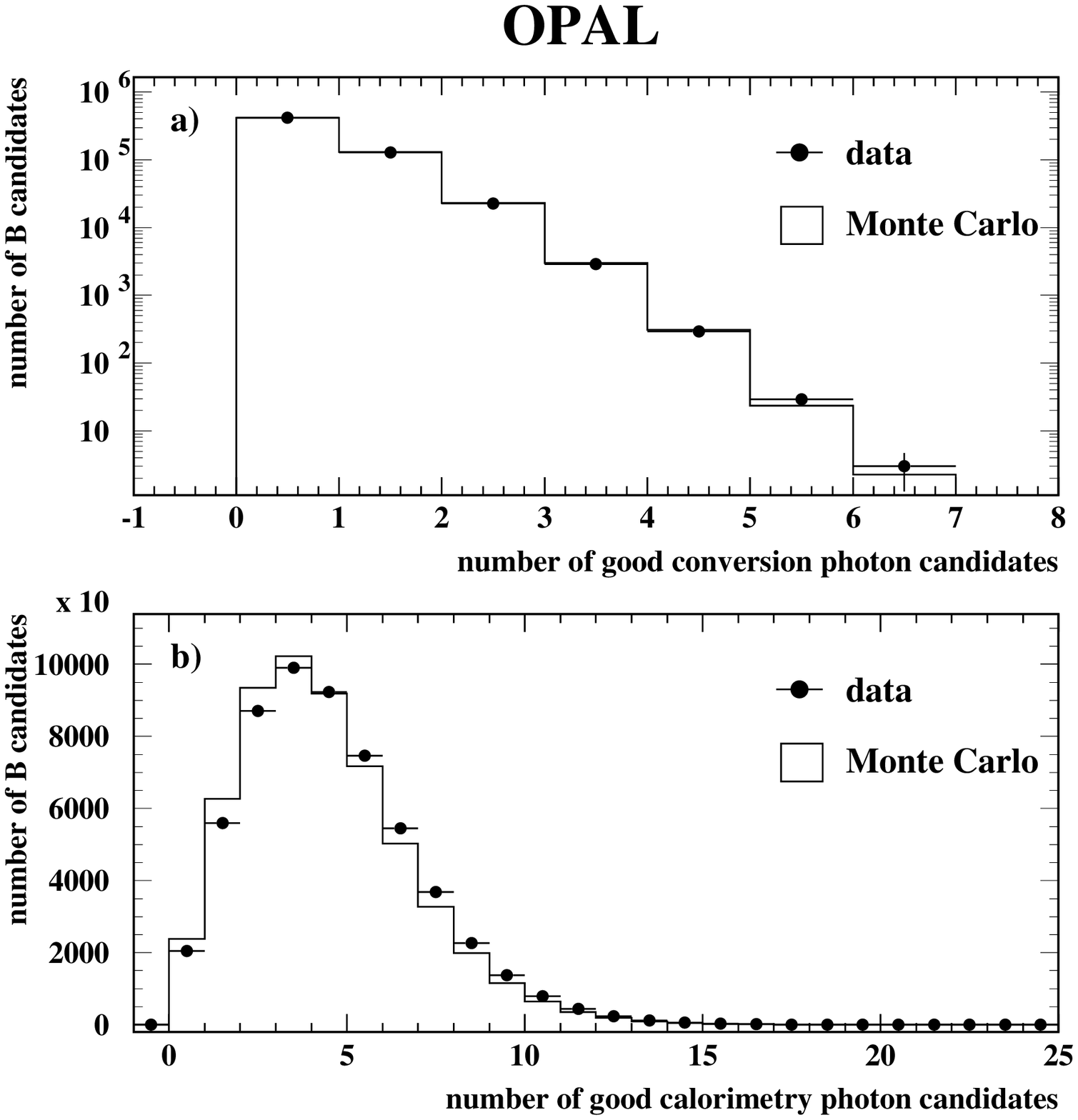}{f:ncand}{a) The number of good conversion photon
  candidates per B candidate observed in data and Monte Carlo. b) The
  number of good calorimeter photon candidates per B candidate
  observed in data and Monte Carlo. For the analysis, the Monte Carlo
  distribution of the latter is reweighted to the data distribution.}

\epostfig{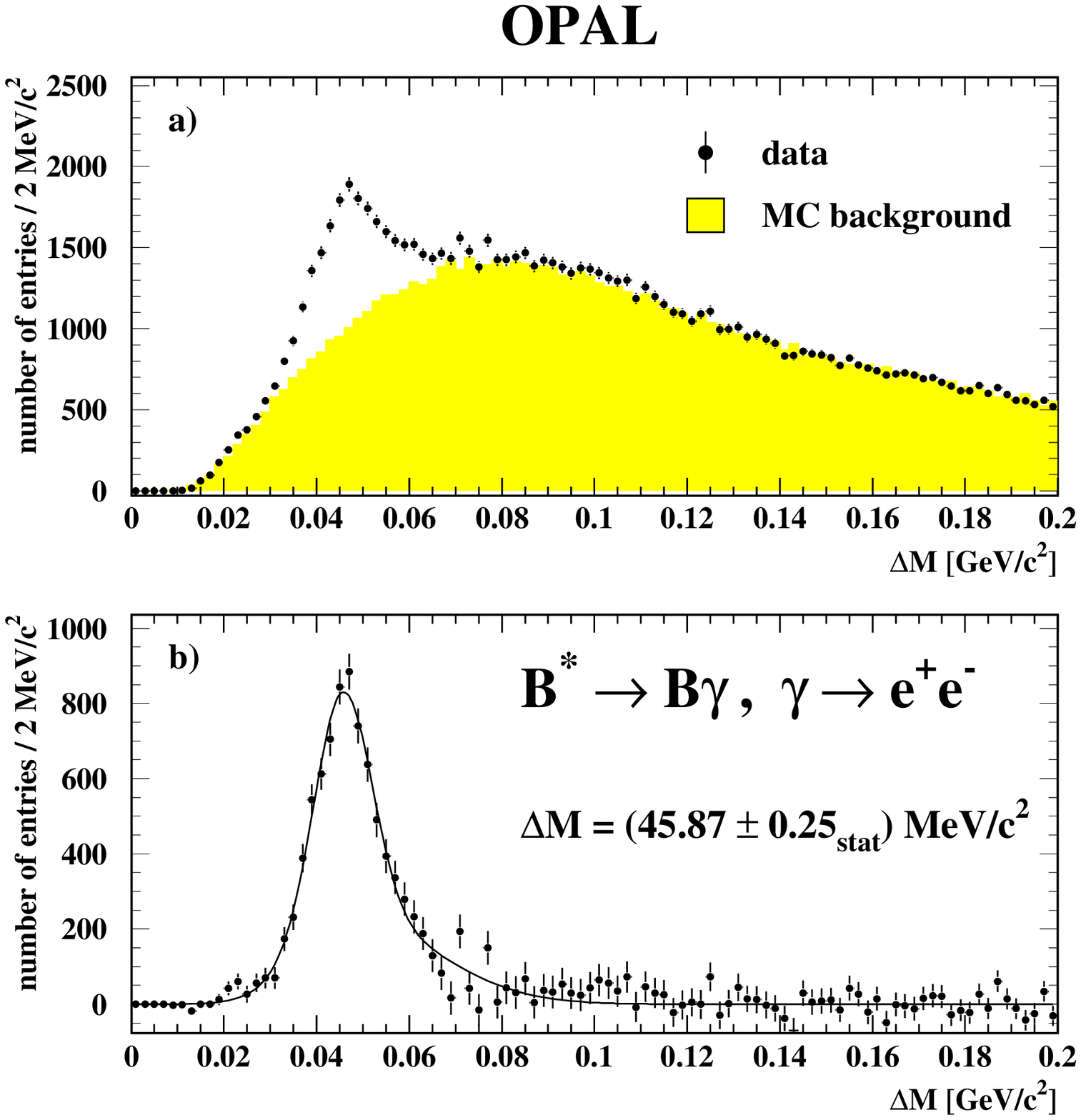}{f:bsconv}{a) The $\Delta M=M_{\rm
    B\gamma}-M_{\rm B}$ mass distribution of the conversion photon
  sample. The background is estimated from Monte Carlo simulation and
  normalised to the data distribution in the sideband region $0.09\;
  \GeVcc < \dmbstar < 0.20\; \GeVcc$. b) The corresponding background
  subtracted signal. The fit function used for the signal is described
  in the text.}

\epostfig{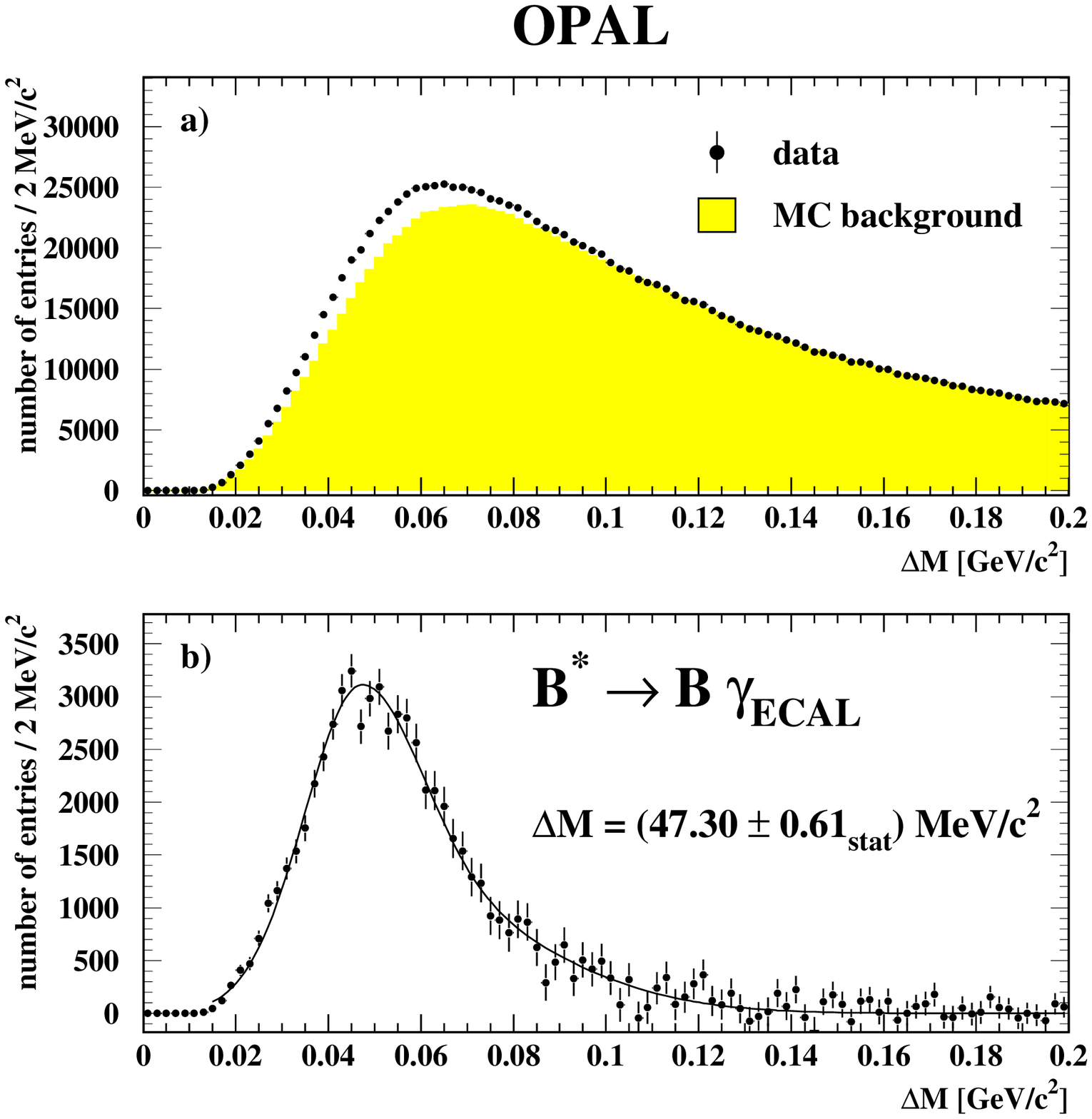}{f:bsecal}{a) The $\dmbstar = M_{\rm
    B\gamma}-M_{\rm B}$ mass distribution of photons reconstructed in
  the electromagnetic calorimeter. The background is estimated from
  the Monte Carlo simulation and normalised to the data distribution
  in the sideband region $0.10\; \GeVcc < \dmbstar < 0.20\; \GeVcc$.
  Although the resolution is poor compared to the conversion photon
  sample, an excess of entries in the data distribution around $46\;
  \MeVcc$ is clearly visible. b) The corresponding background
  subtracted signal. The fit function is described in the text.}

\epostfig{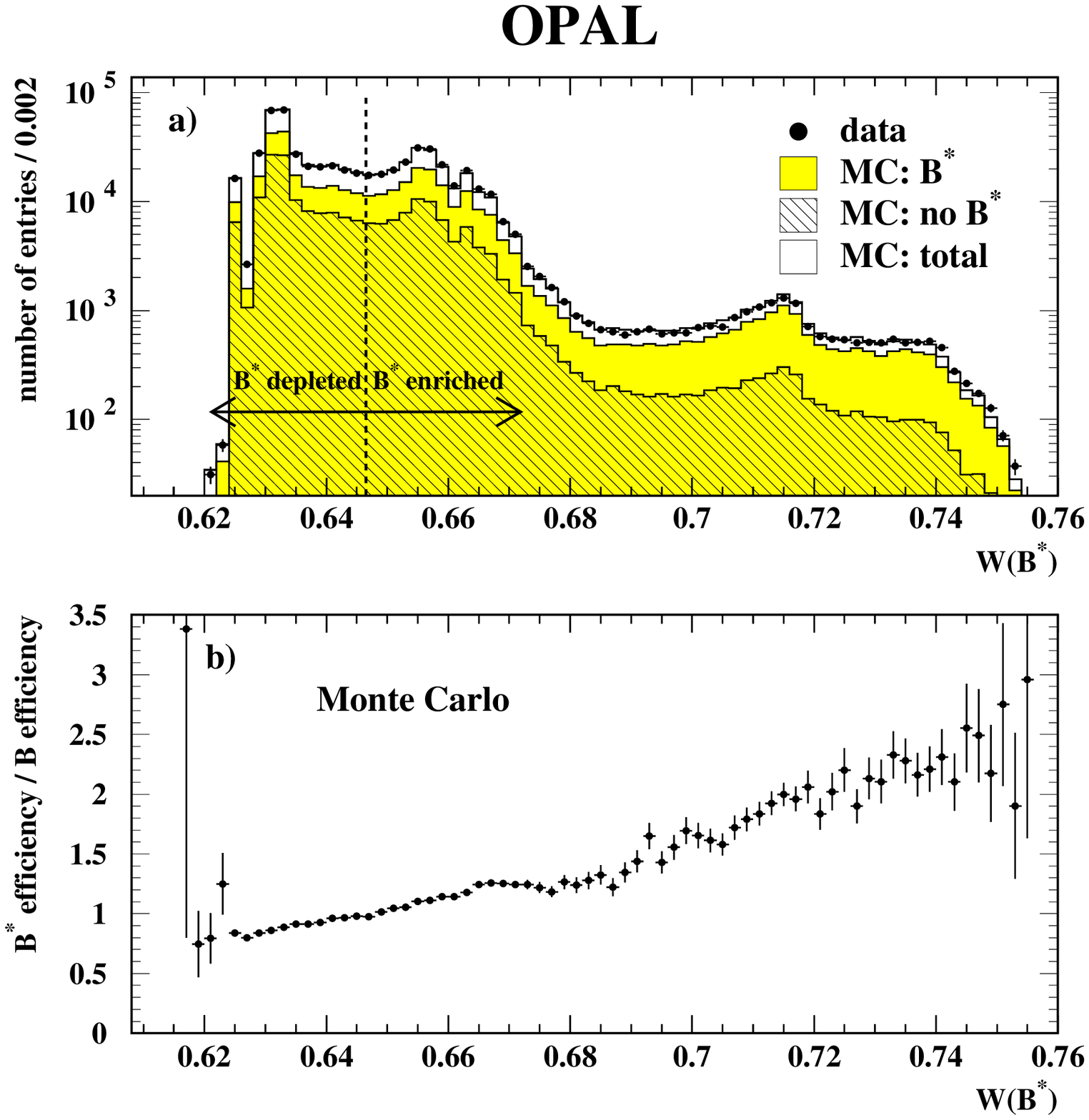}{f:bsprob}{a) The $\bsprob$ distribution for data
  with the corresponding Monte Carlo histograms indicating the number
  of B candidates with a $\bstar$ parent and no $\bstar$ parent. The
  dotted line gives the boundary between the $\bstar$-enriched and
  $\bstar$-depleted samples. b) The ratio of the efficiency to
  reconstruct a B meson with a $\bstar$ parent over the efficiency to
  reconstruct a B meson without a $\bstar$ parent versus the weight
  $\bsprob$ calculated from simulated data.}

\epostfig{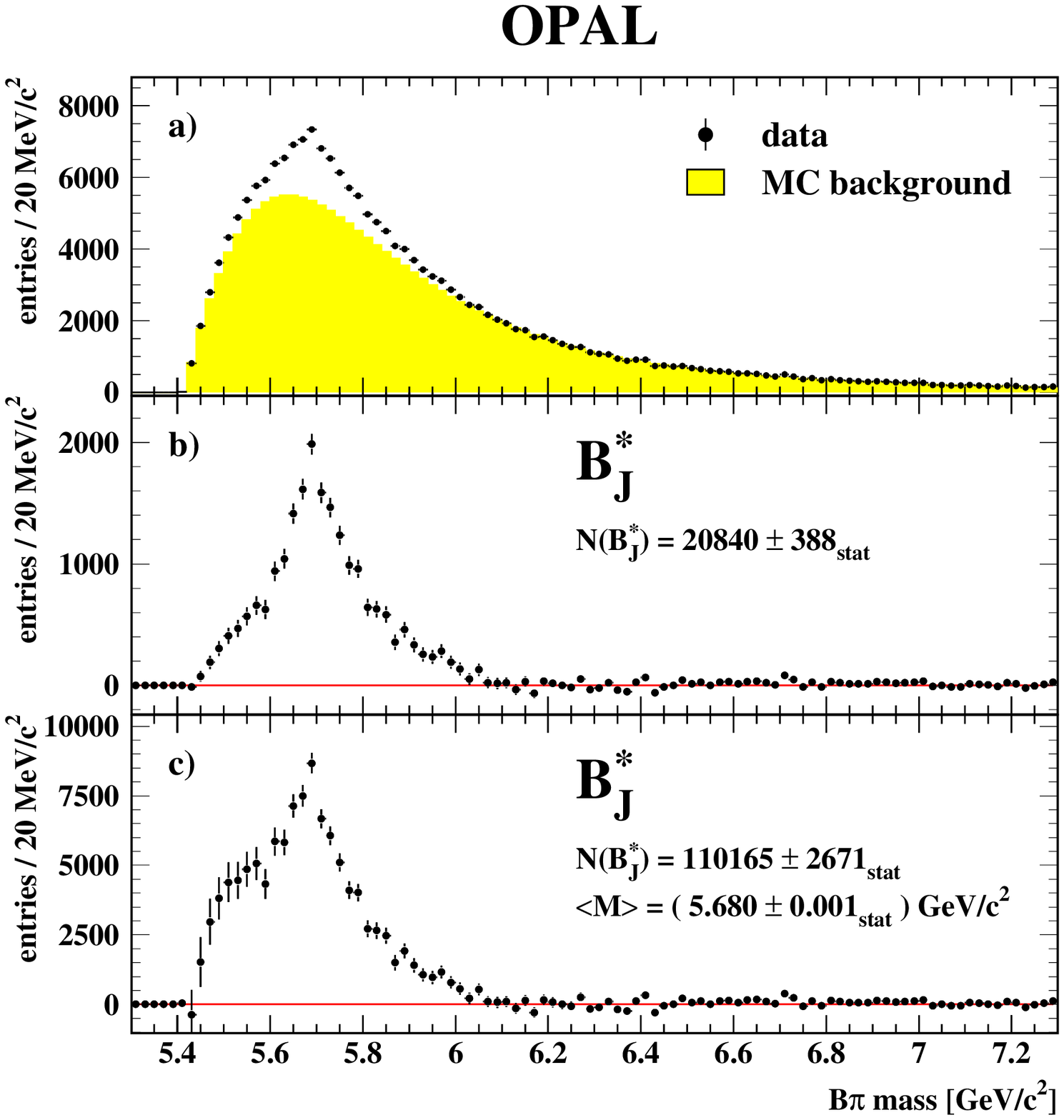}{f:bdsall}{a) The $\bpipm$ mass distribution for
  data. The shaded histogram indicates a fit to the corrected Monte
  Carlo background using a reweighting method described in
  Section~\ref{ss:syspion}. The function used for the fit is described
  in the text. b) The $\bstarj$ signal after subtraction of the
  simulated background. c) The efficiency-corrected $\bstarj$
  signal. The observed structure of the $\bstarj$ signal suggests a
  superposition of several different states. The mass dependent
  efficiency correction has a strong impact on the signal shape at low
  $\bpi$ mass values.}

\epostfig{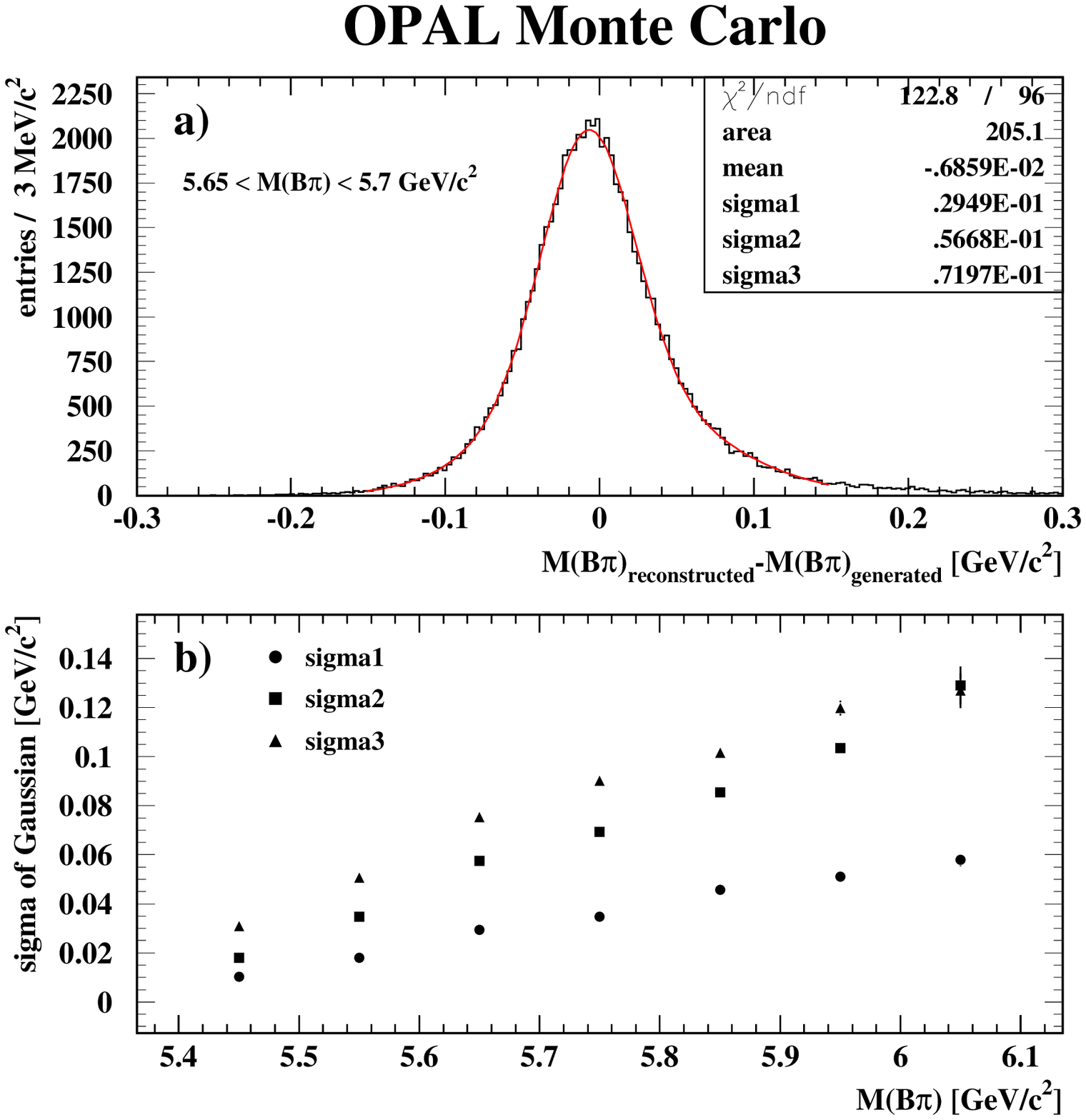}{f:bpireso}{ a) Monte Carlo $M_{\rm B\pi}$
  resolution of $\bstarj$ decaying to $\rm B^{(*)}\pi^\pm$ in the mass
  region $5.65\;\GeVcc < M(\rm B\pi) < 5.70\;\GeVcc$. The fit function
  is the sum of two Gaussians both constrained to the same mean value.
  Sigma1 is the standard deviation of the narrow Gaussian and sigma2
  (sigma3) corresponds to the left (right) standard deviation of the
  asymmetric broad Gaussian. b) The linear dependence of the width of
  the resolution function on $M_{\rm B\pi}$ is shown for each sigma in
  the $\bstarj$ signal region.}


\epostfig{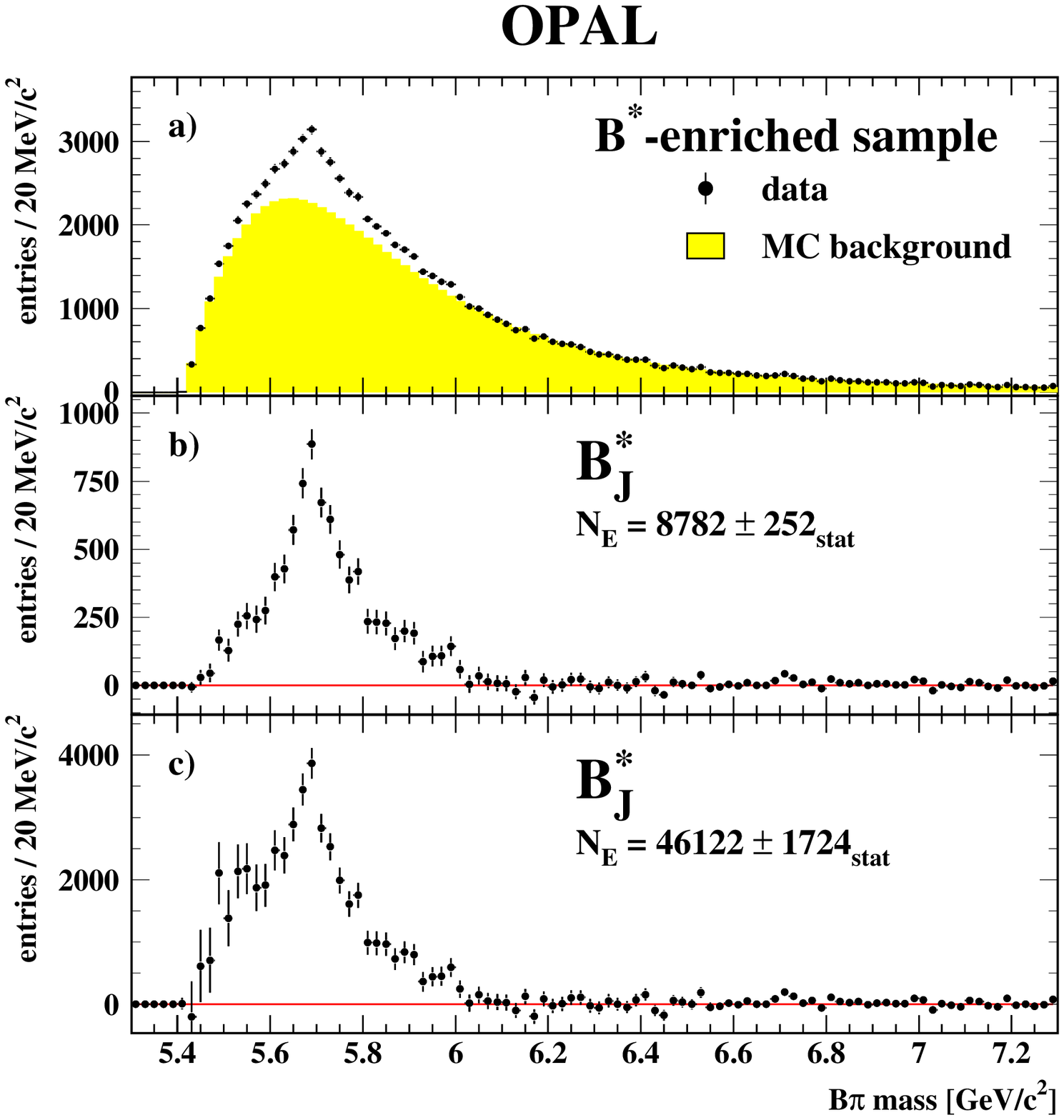}{f:bsenr}{a) The $\bpipm$ mass distribution of the
  sample enriched in the decay $\bstarjtobsb$ in data. The shaded
  histogram indicates a fit to the corrected Monte Carlo background
  using a reweighting method described in Section~\ref{ss:syspion}. b)
  The signal distribution after subtraction of the simulated
  background. c) The efficiency-corrected signal.}

\epostfig{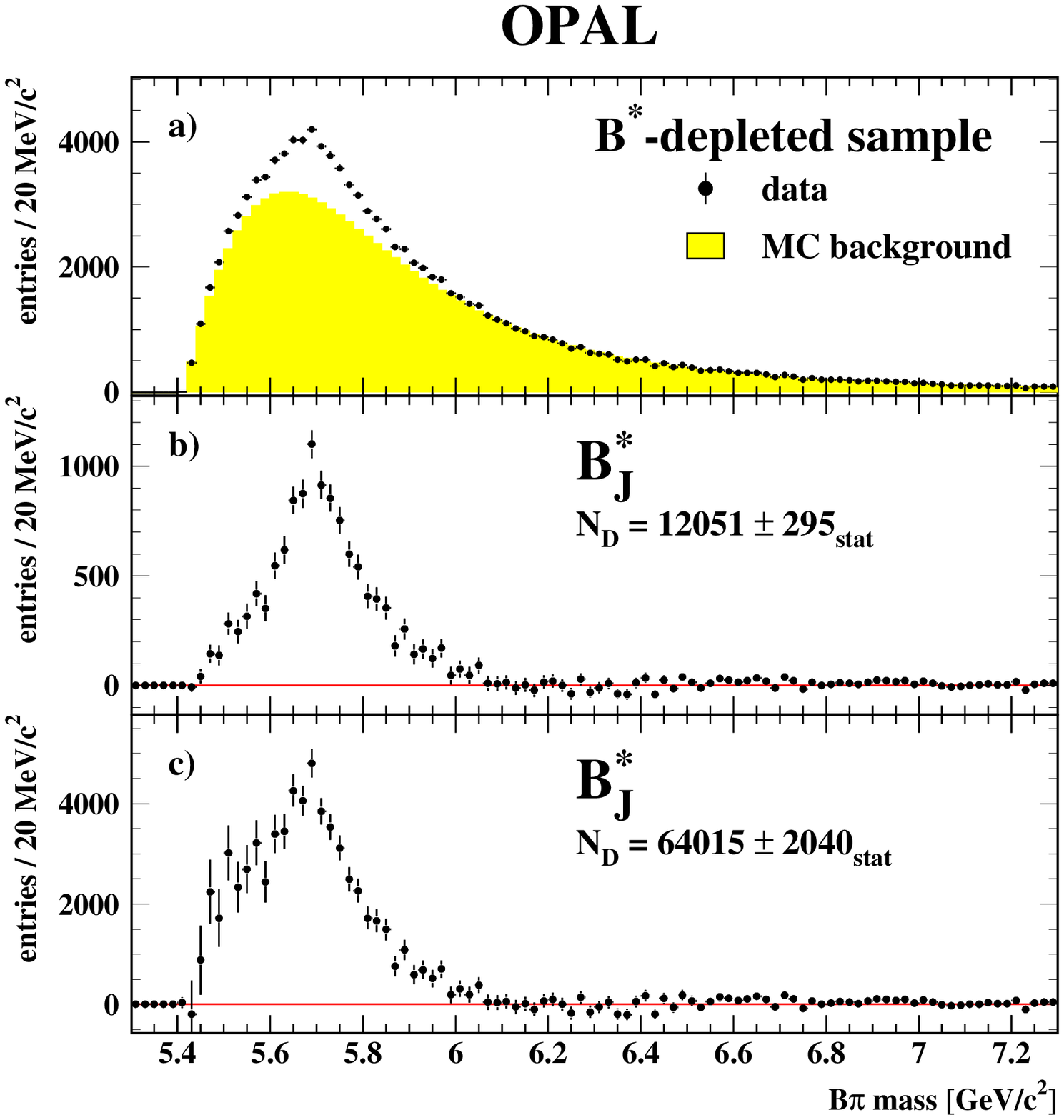}{f:bsdep}{a) The $\bpipm$ mass distribution of the
  sample depleted in the decay $\bstarjtobsb$ in data. The shaded
  histogram indicates a fit to the corrected Monte Carlo background
  using a reweighting method described in Section~\ref{ss:syspion}. b)
  The signal distribution after subtraction of the simulated
  background. c) The efficiency-corrected signal.}

\epostfig{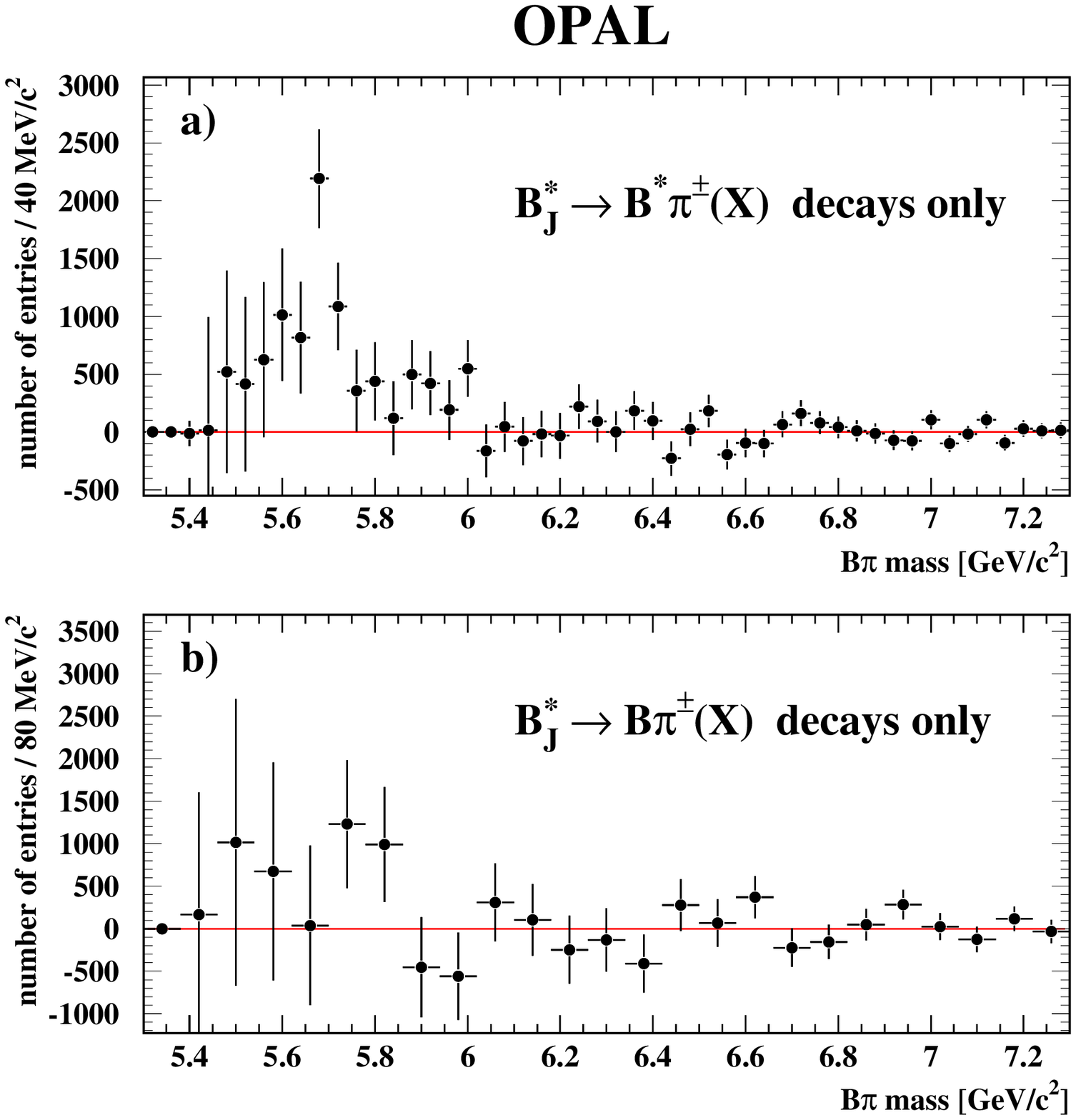}{f:pure}{a) The efficiency-corrected $\bpipm$ mass
  distribution of $\bstarjtobsb$ transitions seen in data. A clear
  peak is visible at $5.7\;\GeVcc$. The structure is unlikely to stem
  from a single state. b) The efficiency-corrected $\bpipm$ mass
  distribution of $\bstarjtobb$ transitions seen in data. A
  $2.2\sigma$ excess is observed around $5.8\;\GeVcc$.}

\epostfig{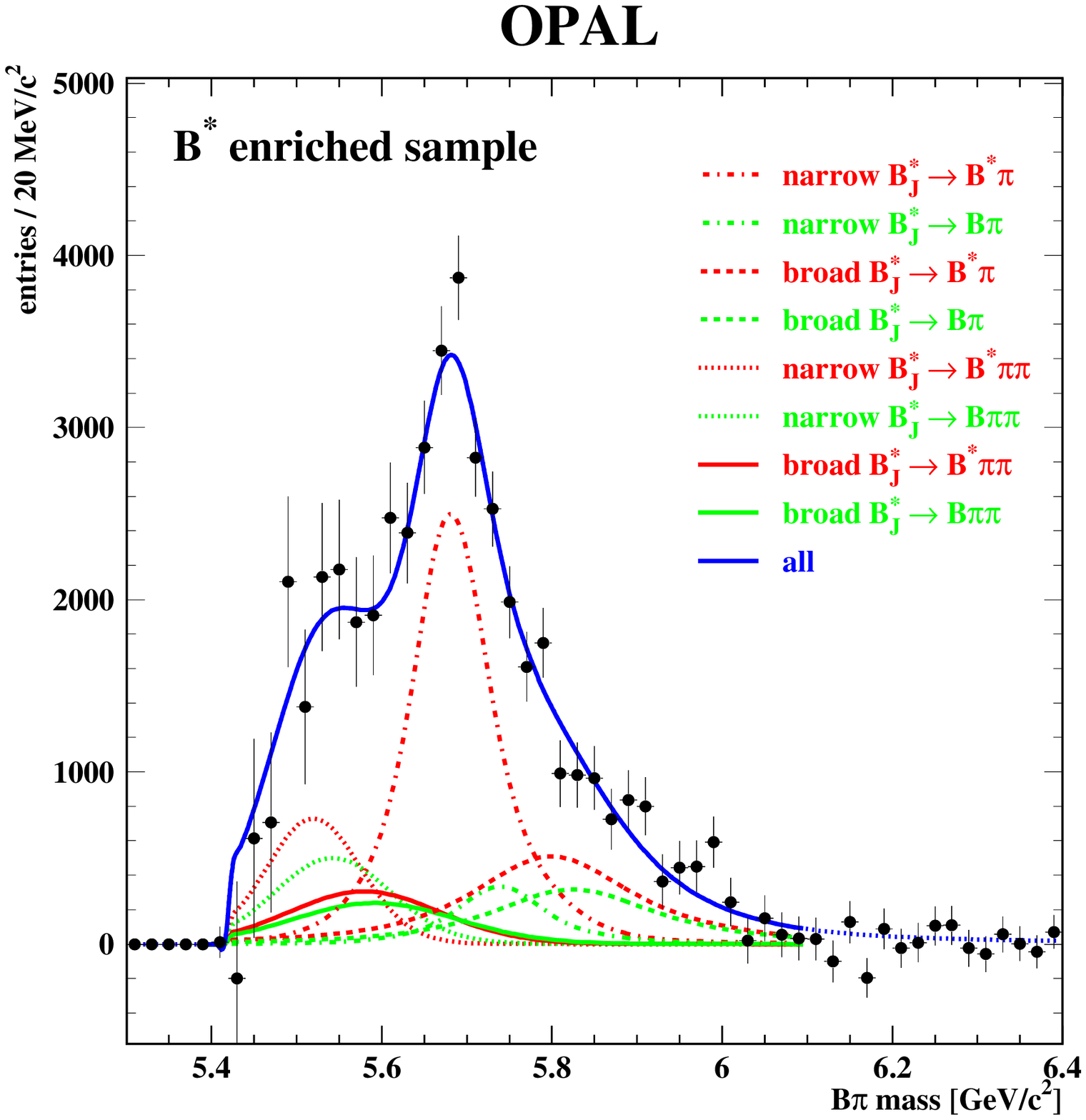}{f:bsenrfit}{ Simultaneous fit to the $\bpi$
  mass distributions of the $\bstarj$ samples enriched or depleted in
  $\bstarj\rightarrow \bstar \pi^\pm (\pi)$ decays (see also
  Figure~\ref{f:bsdepfit}). The fit results of the decays of the broad
  and narrow $\bstarj$ for transitions via one and two pions are
  presented separately.}

\epostfig{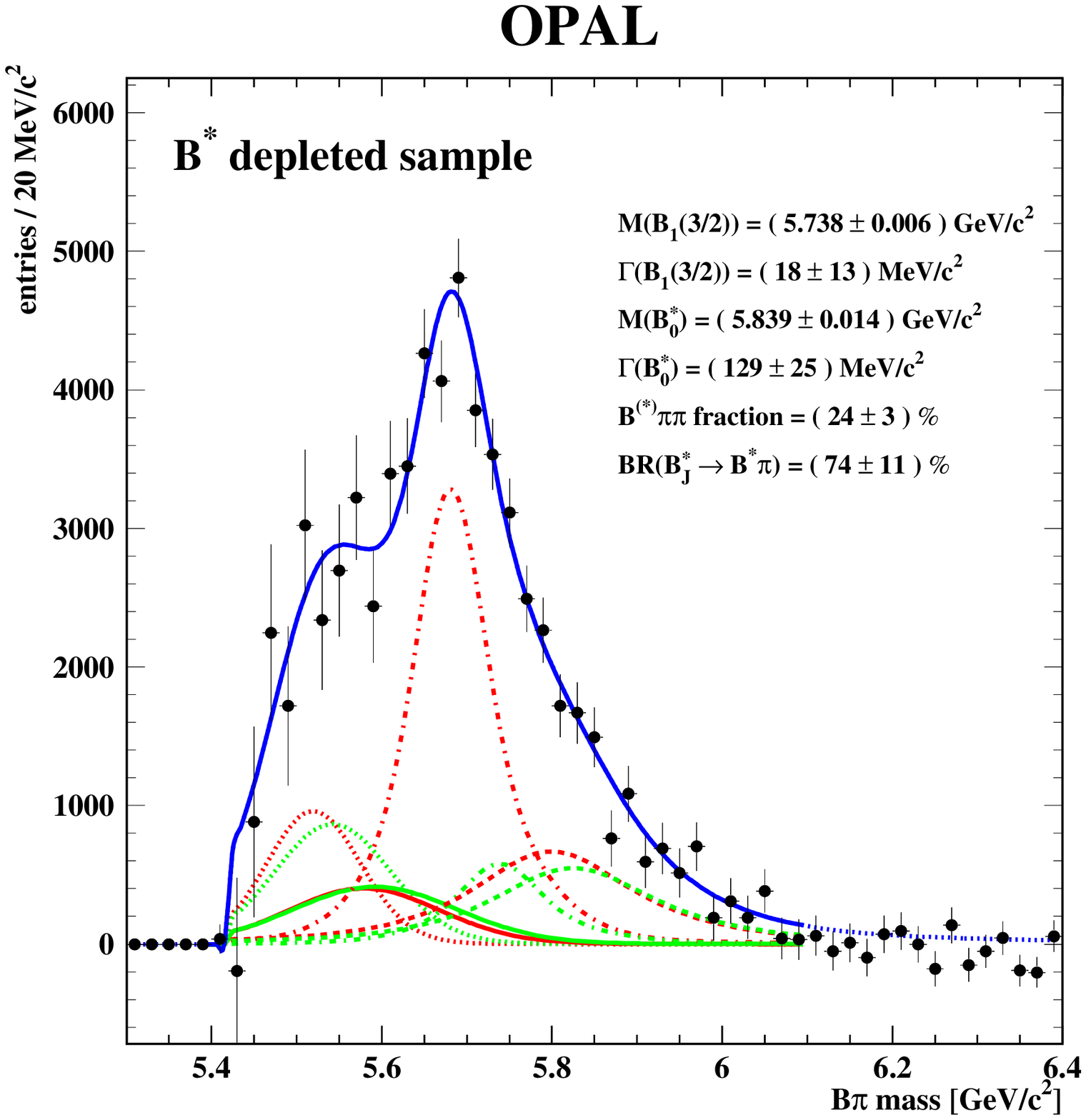}{f:bsdepfit}{ Simultaneous fit to the $\bpi$
  mass distribution of the $\bstarj$ samples enriched or depleted in
  $\bstarj\rightarrow \bstar \pi^\pm (\pi)$ decays. The fit results of
  the decays of the broad and narrow $\bstarj$ for transitions via one
  and two pions are presented separately. In comparison to
  Figure~\ref{f:bsenrfit} the fraction of $\bstarj$ decays to $\bstar$
  is reduced \wrt\ $\bstarj$ decays to $\rm B$ (e.g. compare the light
  ($\bstarj\rightarrow \bstar \pi\pi$) and the dark
  ($\bstarj\rightarrow \rm B \pi\pi$) solid lines in
  Figures~\ref{f:bsenrfit} and \ref{f:bsdepfit}).}

\epostfig{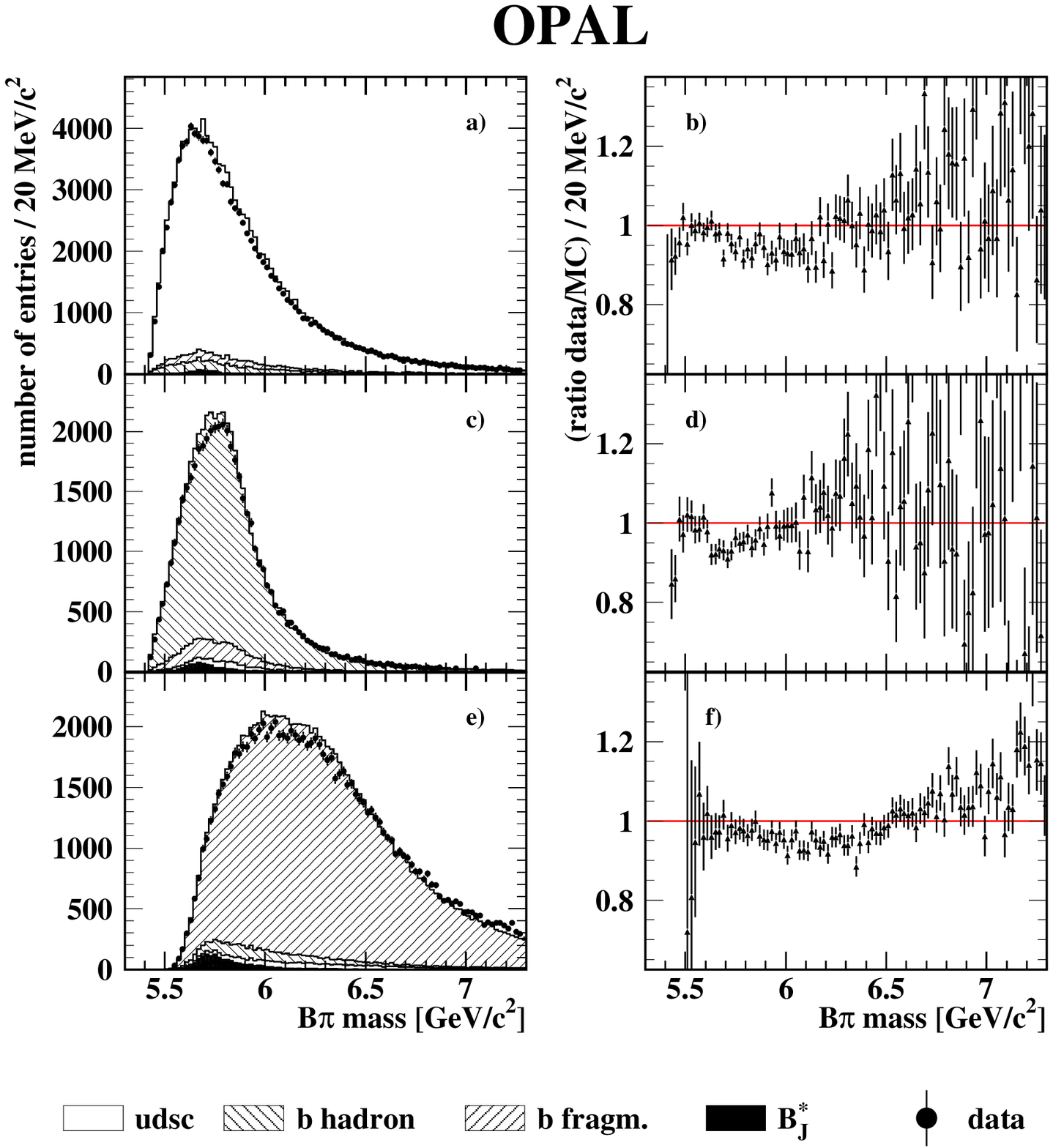}{f:bgcorr}{ The $\rm B\pi^\pm$ mass distributions
  of data and Monte Carlo for each of the three test samples (left
  side) and the corresponding bin-by-bin ratio of the mass
  distributions (right side). a)+b) Light and charm quark sample,
  c)+d) b hadron decay sample, e)+f) b fragmentation sample. }

\end{document}